\documentclass[11pt]{article}

\usepackage{aas_macros,amsmath,amssymb,comment,cite,esint,graphicx,mathtools}
\usepackage[margin=.8in,letterpaper]{geometry}
\usepackage[colorlinks=true]{hyperref}
\usepackage[affil-it]{authblk}
\usepackage{subcaption}
\usepackage[utf8]{inputenc}
\usepackage{mathrsfs}
\usepackage{appendix}
\usepackage{amssymb}
\usepackage{float}
\usepackage{color}
\usepackage{cite}
\usepackage{hyperref}
\hypersetup{pageanchor=false}
\usepackage{indentfirst}
\usepackage{url}
\usepackage{float}
\usepackage{caption}
\usepackage[numbers,square,comma,sort&compress,merge]{natbib}
\usepackage{esint}
\usepackage{overpic}
\usepackage{graphicx}
\usepackage{epsf,amsmath,bbold,amsfonts,stmaryrd}
\usepackage{textcomp}
\usepackage{ulem}
\usepackage{tikz}

\numberwithin{equation}{section}
\let\originalleft\left
\let\originalright\right
\renewcommand{\left}{\mathopen{}\mathclose\bgroup\originalleft}
\renewcommand{\right}{\aftergroup\egroup\originalright}
\mathcode`\*="8000
{\catcode`\*=\active\gdef*{\mathclose{}\,\mathopen{}}}

\newcommand{\be}{\begin{equation}}
\newcommand{\ee}{\end{equation}}
\newcommand{\bea}{\setlength\arraycolsep{2pt} \begin{eqnarray}}
\newcommand{\eea}{\end{eqnarray}}

\begin{document}
\title{Energy extraction from a rotating black hole via magnetic reconnection: Bumblebee gravity}

\author{
Ho-Yun YuChih$^{2}$~,
Ye Shen$^{1}$\thanks{E-mail: shenye199594@stu.pku.edu.cn},~
}
\date{}

\maketitle
\vspace{-10mm}

\begin{center}
	{\it	
      $^1$College of Physics and Information Engineering, Fuzhou University, \\
		Fuzhou 350108, Fujian, P. R. China \\
      $^2$School of Physics, Peking University, No.5 Yiheyuan Rd, Beijing
		100871, P.R. China\\\vspace{4mm}
}
\end{center}

\vspace{8mm}

\begin{abstract}
	\vspace{5mm}

Many efforts were made in order to better understand the energy extraction via magnetic reconnection from a rotating black hole, following the work of Comisso and Asenjo in 2021. We also tried to make some progress in our previous works, in which we discussed differences between bulk plasma with different streamlines and also defined the covering factor as an internal property of an accretion system to quantify its capability on extracting energy via magnetic reconnection from its central black hole. In this study, we aim to explore this topic within the framework of a Kerr–Sen-like spacetime induced from Bumblebee gravity, which, among various alternative theories of gravity beyond pure Einstein gravity, stands out as a promising candidate for explaining certain high-energy astrophysical phenomena. More specifically, we would like to analyze the influence of the rate of Lorentz symmetry breaking and the Bumblebee charge, the two additional parameters in Bumblebee gravity except for the black hole mass and spin, on the energy extraction via magnetic reconnection. By analyzing the allowed regions for energy extraction and the variations of covering factor, we find that energy extraction becomes more likely to succeed and tends to occur closer to the central region when the spacetime carries bigger rate of Lorentz symmetry breaking and Bumblebee charge. Furthermore, our results indicate that the most favorable spacetime configuration for energy extraction via magnetic reconnection, when the extractable energy of the central black hole is determined, corresponds to the scenario in which the cosmic censorship hypothesis is marginally not violated.

\end{abstract}

\maketitle

\newpage
\baselineskip 18pt

\section{Introduction}
\label{sec:intro}

With increasing support from numerical simulations, magnetic reconnection, in which electromagnetic energy is efficiently converted into the kinetic or thermal energy of plasma in a magnetofluid, has been shown to occur frequently in the accretion flows near black holes \cite{Yuan2009,Yuan2024-1,Yuan2024-2,Nathanail:2024efu}. It is speculated that the infrared and X-ray flares, observed from the supermassive black hole Sagittarius A*, are generated by magnetic reconnection processes \cite{Ripperda2020,Aimar2023,Sen:2025ytv}. Based on these, extracting energy from a rotating black hole via magnetic reconnection occurring in the ergosphere has been proposed as a more realizable mechanism than the classical Penrose process \cite{Penrose,Wald:1974kya}. The possibility of this conception was first investigated by Koide and Arai \cite{KA2008}. In their work, magnetic reconnection was set to occur in the bulk plasma flows circularly orbiting a Kerr black hole. Reference~\cite{KA2008} further demonstrated that in an astrophysical scenario, energy extraction from a Kerr black hole is more feasible via magnetic reconnection than via the fission of particles as described by Penrose process.

Comisso and Asenjo enriched the framework and introduced the plasmoid dynamics to describe the magnetic reconnection processes on the equatorial plane near a rotating black hole \cite{CA2021}. In the region extremely close to the event horizon, the large-scale configuration of magnetic field is predominantly determined by the gravitational effect. Analytically, antiparallel magnetic field lines would form adjacent to the equatorial plane due to the symmetries of spacetime \cite{Split-Mono,Gralla:2014yja}, which was already supported by numerical simulations \cite{Sasha2011,McKinney:2006sc,Vos:2024loa}. The flip in magnetic field direction explains both the formation of equatorial current sheet and the expectation of magnetic reconnection occurring on the equatorial plane. The equilibrium of antiparallel of magnetic fields would be broken by plasmoid instability, triggering fast magnetic reconnection and expelling a pair of plasmoids from the reconnection point. \cite{Comisso:2016pyg,Comisso:2017arh}. Upon the plasmoid instability occurring in the ergoregion, a plasmoid carrying negative energy-at-infinity density would be generated so that the energy extraction from the rotating black hole is realized. The process proposed in Ref.~\cite{CA2021}, as briefly summarized above, is now widely referred to be the Comisso–Asenjo process. This process has also been investigated in various stationary, axisymmetric spacetimes other than the pure Einstein gravity 
\cite{Carleo:2022qlv,Khodadi:2022dff,Wei:2022jbi,Liu:2022qnr,Wang:2022qmg,Shaymatov:2023dtt,ZhangShaoJun2024,Long:2024tws,Zeng:2025vjt,Zeng:2025olq}.

Theoretically, a charged and rotating black hole was proposed under the scheme of heterotic string theory, which is widely known as the Kerr-Sen black hole \cite{Sen:1992ua}. Meanwhile, studies on Lorentz symmetry breaking in high-energy limit gain considerable attention in recent years \cite{Li:2023wlo,Khodadi:2023yiw,Panotopoulos:2024jtn}. Both are considered primary candidates for probing quantum gravity. A Kerr–Sen–like solution has been derived from the Einstein-Bumblebee action, in which two additional parameters, corresponding to the rate of Lorentz symmetry breaking $l$ and the charge in Kerr-Sen black hole $b$, are spontaneously included \cite{Casana:2017jkc,Jha:2020pvk,Ding:2019mal,Xu:2022frb}. The constraints (or estimations) of these two parameters were carefully made from observations of supermassive black holes \cite{Zhu:2024qcm,Jha:2021eww,Wang:2021gtd,Islam:2024sph,Gu:2022grg}.

In our previous work, taking the Comisso–Asenjo process as the basis, we investigated the influence of the bulk plasma streamlines, orientation angles, and parameters in reconnection models \cite{Work0,Work1,Work2}. As a preliminary attempt to quantify the capability of an accretion system on extracting energy from its central black hole via magnetic reconnection, we introduced the concept of the covering factor. In this work, we aim to extend our discussions to the Kerr–Sen–like black hole arising from the Einstein-Bumblebee gravity. 

To set the stage, we will revisit the calculations in Comisso–Asenjo process and examine some basic properties of the Kerr–Sen–like black hole for preparation. The allowed regions for energy extraction in $a–r_0$ and $\xi_{\rm B}–r_0$ parameter planes for various values of $(l,b)$ will be exhibited, where $a$, $r_0$ and $\xi_B$ represent the black hole spin, the radial position of reconnection point and the orientation angle, respectively. Six sets of $(l,b)$ will be studied as examples. The influences of $l$ and $b$ on energy extraction will then be discussed. Furthermore, we will present the distributions of covering factor in $l–b$ parameter planes for various black hole spins and bulk plasma streamlines. From these, one can easily figure out how the capability of an accretion system to energy extraction via magnetic reconnection varies with the existence of Lorentz symmetry breaking and Bumblebee charge. To further analyze the impact of these two parameters, we will also plot the variation of covering factor along the line of $l=0.3$, the line of $b=0.3$ and the line with fixed irreducible mass (which will be called the contour line of irreducible mass hereafter). Our results indicate that a higher rate of Lorentz symmetry breaking or Bumblebee charge will strengthen the capability of an accretion system on extracting energy via magnetic reconnection from its central black hole. Additionally, we will show that the most favorable condition for energy extraction occurs when the cosmic censorship hypothesis is marginally not violated, when the extractable energy of the central black hole remains fixed.

The remainder of this paper is organized as follows. In Sect.~\ref{sec:ca}, we will revisit the basic calculations involved in the Comisso–Asenjo process. In Sect.~\ref{sec:Kerr-Sen}, the Kerr–Sen–like metric coming from Einstein-Bumblebee gravity and its geodesics will be introduced. Our main results will be presented in Sect.~\ref{sec:res}. In Sect.~\ref{sec:prop}, we will examine several properties of the Kerr–Sen-like spacetime and its characteristic radii, as preparations for the subsequent analyses. In Sect.~\ref{sec:allow}, the allowed regions for energy extraction in different parameter planes will be shown. Detailed analyses of the covering factor are shown in Sect.~\ref{sec:covering}, from which we will try to investigate how the energy extraction is affected by the rate of Lorentz symmetry breaking and the Bumblebee charge. We will briefly summarize our work in Sect.~\ref{sec:sum}. Throughout this work, we adopt the natural units  with $G=M=c=1$ without loss of generality. Unless otherwise stated, the notation and conventions follow those of Refs.~\cite{Work1,Work2}.

\section{Comisso–Asenjo process}
\label{sec:ca}

We revisit some basic concepts introduced in Ref.~\cite{Work0,Work1}, which are necessary to quantify the energy extraction via magnetic reconnection from a rotating black hole. For the physical process occurring in a stationary, axisymmetric spacetime, it is effective to represent equations in the 3+1 formalism \cite{MacDonald:1982zz}, where the line element can be expressed as:
\begin{equation}
    ds^2=g_{\mu\nu}dx^{\mu}dx^{\nu}=-\alpha^2dt^2+\sum_{i=1}^3\left(h_idx^i-\alpha\beta^idt\right)^2
    \label{eq:line}
\end{equation}
where $\alpha$ is the lapse function and $\beta^i=h_i\omega^i/\alpha$ is the shift vector, with $\omega^i$ being the velocity of frame dragging and $h_i$ being the scale factor. In this study, we are interested in the Kerr–Sen–like metric, which can be expressed in 3+1 form as well, with all the components slightly different from those in Kerr metric. We will introduce the Kerr–Sen–like metric in detail in Sect.~\ref{sec:Kerr-Sen}. Upon the normal tetrad of zero-angular-momentum-observers (ZAMOs) $\hat{e}^{\mu}_{(\nu)}$, the fluid's rest frame, in which the magnetic reconnection will be quantified, could be defined via the normal tetrad as:
\begin{equation}
    \begin{split}
        e_{[0]}^{\mu}&=\hat{\gamma}_s \left[\hat{e}_{(t)}^{\mu}+\hat{v}_s^{(r)}\hat{e}_{(r)}^{\mu}+\hat{v}_s^{(\phi)}\hat{e}_{(\phi)}^{\mu}\right]; \\
        e_{[1]}^{\mu}&=\frac{1}{\hat{v}_s}\left[\hat{v}_s^{(\phi)}\hat{e}_{(r)}^{\mu}-\hat{v}_s^{(r)}\hat{e}_{(\phi)}^{\mu}\right],~~
        e_{[2]}^{\mu}=\hat{e}_{(\theta)}^{\mu}, \\
        e_{[3]}^{\mu}&=\hat{\gamma}_s\left[\hat{v}_s\hat{e}_{(t)}^{\mu}+\frac{\hat{v}_s^{(r)}}{\hat{v}_s}\hat{e}_{(r)}^{\mu}+\frac{\hat{v}_s^{(\phi)}}{\hat{v}_s}\hat{e}_{(\phi)}^{\mu}\right]
    \end{split}
    \label{eq:fluid}
\end{equation}
where $\hat{v}_s=\sqrt{\left(\hat{v}_s^{(r)}\right)^2+\left(\hat{v}_s^{(\phi)}\right)^2}$ is the speed of fluid, with $\hat{v}_s^{(r)}$ and $\hat{v}_s^{(\phi)}$ representing the components of 3-velocity, and $\hat{\gamma}_s$ is the Lorentz factor, in the view of ZAMOs. It is clear that $e_{[1]}$ and $e_{[3]}$ are orthogonal and parallel to the fluid velocity, respectively.

The bulk plasma is assumed to move geodesically, whose 4-velocity $U^{\mu}$ could be projected onto the normal tetrad of ZAMOs as:
\begin{equation}
    U^{\mu}\hat{e}^{(a)}_{\mu}=\hat{\gamma}_s\left\{1,\hat{v}_s^{(r)},0,\hat{v}_s^{(\phi)}\right\}=
    \left\{\frac{E-\omega^{\phi}L}{\alpha},h_rU^r,0,\frac{L}{h_{\phi}}\right\}
    \label{eq:U}
\end{equation}
where $E$ and $L$ are two conserved quantities along geodesics in stationary axisymmetric spacetime. Similar to Ref.~\cite{Work1,Work2}, we take two kinds of streamlines of the bulk plasma into consideration. The first is called the combined streamline, for which the bulk plasma flows circularly outside the innermost stable circular orbit (ISCO) and plunges to the central black hole from ISCO. The second is called the circular streamline, for which the bulk plasma flows circularly all the way down to the limit given by photon sphere. More information, especially the details about the circular geodesics and plunging geodesic in Kerr-Sen spacetime will be introduced in Sect.~\ref{sec:geodesic}. Our focus on the equatorial plane is not merely for simplification. As we know, the electromagnetic field extremely near the black hole could be treated using force-free electrodynamics \cite{McKinney:2006sc,Gralla:2014yja} when the magnetization is sufficiently high \cite{McKinney:2006sc,Gralla:2014yja}. With this treatment, the global magnetic field configuration near a rotating black hole resembles the so-called split-monopole.  \cite{Split-Mono,McKinney:2006sc,Sasha2011,Gralla:2014yja,Vos:2024loa}. In this case, a broad equatorial current sheet forms. Namely, antiparallel magnetic field lines, which is the initial condition of magnetic reconnection, appears naturally on two sides of the equatorial plane.

The the local structure of antiparallel magnetic field lines is unstable due to the plasmoid instability \cite{Comisso:2016pyg,Comisso:2017arh}. Under tiny perturbation, multiple fast magnetic reconnection processes occur, leading to the generation of plasmoid pairs. The 4-velocities of ejected plasmoids are given by
\begin{equation}
    u^{\mu}_{\pm}=\gamma_{\rm out}\left[e_{[0],0}^{\mu}\pm 
    v_{\rm out}\left(\cos\xi_{B}e_{[3],0}^{\mu}+\sin\xi_{B}e_{[1],0}^{\mu}\right)\right]
    \label{eq:u_out}
\end{equation}
where the "$\pm$" indicates that two generated plasmoids are ejected in opposite directions. We adopt a theoretical model describing the fast magnetic reconnection, in which the outflow speed obeys \cite{Liu2017}:
\begin{equation}
    v_{\rm out}\simeq \sqrt{\frac{\left(1-{\sf g}^2\right)^3\sigma_0}{\left(1+{\sf g}^2\right)^2+\left(1-{\sf g}^2\right)^3\sigma_0}}
    \label{eq:v_out}
\end{equation}
Here, $\sigma_0$ is the local magnetization on reconnection point\footnote{It is worth emphasizing that the "local" here is mentioned with respect to the global scale that is generally $\sim r_g$. Similarly, the "reconnection point" can be regarded as a point only with respect to the global scale. If we stare at the reconnection point specifically, the region within which the magnetic reconnection occurs, $\sigma_0$ should be the magnetization upstream. Please revisit Ref.~\cite{Work0,Work1} or Ref.~\cite{Liu2017} for detailed explanations.}. While $g$, referred to as the geometric index, is defined as the thickness-to-length ratio of the current sheet. We choose ${\sf g}=0.49$ so that the reconnection rate is maximized in the limit of high magnetization \cite{Work0}. The orientation angle $\xi_B$ is treated as a free parameter ranging from $-\pi/2$ to $\pi/2$ (see Fig.~1 in Ref.~\cite{Work1}). 

The energy extraction efficiency is defined as \cite{CA2021}
 \begin{equation}
    \eta=\frac{\epsilon_+}{\epsilon_++\epsilon_-}
    \label{eq:eta}
\end{equation}
where $\epsilon_{\pm}$ are the energy-at-infinity per unit enthalpy of the two ejected plasmoids, which satisfy \cite{CA2021,Work0}:
\begin{equation}
    \begin{split}
        \epsilon_{\pm}=&-u_t^{\pm}-\frac{\Tilde{p}_{\pm}}{u^t_{\pm}} \\
        =&\alpha\hat{\gamma}_s\gamma_{\rm out}\left[\left(1+\beta^{\phi}\hat{v}_s^{(\phi)}\right)\pm 
        v_{\rm out}\left(\hat{v}_s+\beta^{\phi}\frac{\hat{v}_s^{(\phi)}}{\hat{v}_s}\right)\cos\xi_B\mp
        v_{\rm out}\beta^{\phi}\frac{\hat{v}_s^{(r)}}{\hat{\gamma}_s\hat{v}_s}\sin\xi_B\right] \\
        &-\frac{\alpha\Tilde{p}}{\hat{\gamma}_s\gamma_{\rm out}\left(1\pm \hat{v}_s v_{\rm out}\cos\xi_B\right)}
    \end{split}
    \label{eq:epsilon}
\end{equation}
The pressure per unit enthalpy $\Tilde{p}$ is fixed at $1/4$ \cite{Lyubarsky2006,Comisso:2014nva} in the equation above. The expression in Eq.~\eqref{eq:epsilon} is derived under the assumption that the magnetic energy in plasma is efficiently converted into the kinetic energy of the generated plasmoids in magnetic reconnection. Energy extraction from the central black hole succeeds whenever a plasmoid with negative $\epsilon_-$ is generated. Based on Eq.~\eqref{eq:epsilon}, we define the best orientation angle for energy extraction \cite{Work1}:
\begin{equation}
    \xi_{B,{\rm m}}=\arctan\left(-\frac{\beta^{\phi}\hat{v}_s^{(r)}}{\hat{\gamma}_s\hat{v}_s^2+\beta^{\phi}\hat{\gamma}_s\hat{v}_s^{(\phi)}}\right)
    \label{eq:xi_m}
\end{equation}
which maximizes the energy extraction efficiency in high magnetization limit. For circularly flowing bulk plasma, we have $\xi_{B,{\rm m}}=0$. For plunging bulk plasma, the condition $\xi_{B,{\rm m}}>0$ holds. A positive best orientation angle means the plasmoid with $\epsilon_+$ is ejected more radially outward (as can be seen from Eq.~\eqref{eq:u_out}).

\section{Kerr–Sen–like spacetime}
\label{sec:Kerr-Sen}

\subsection{Metric}
\label{sec:metric}

In Boyer–Lindquist coordinates $(t,r,\theta,\phi)$, the line element of the Kerr–Sen–like metric can be written as \cite{Jha:2020pvk,Ding:2019mal}:
\begin{equation}
    \begin{aligned}
        ds^2=& g_{\mu\nu}dx^{\mu}dx^{\nu} \\
        =&-\left(1-\frac{2r}{\Sigma}\right)dt^2+(1+l)\frac{\Sigma}{\Delta}dr^2+\Sigma d\theta^2+\frac{A\sin^2\theta}{\Sigma}d\phi^2
        -\frac{4\Tilde{a}r\sin^2\theta}{\Sigma}dtd\phi
    \end{aligned}
    \label{eq:metric}
\end{equation}
where\footnote{A more widely used notation is $\Delta=\frac{r(r+b)-2r}{1+l}+a^2$.}
\begin{equation}
    \Sigma=r(r+b)+\Tilde{a}^2\cos^2\theta~~,~~\Delta=r(r+b)-2r+\Tilde{a}^2~~,~~A=\left[r(r+b)+\Tilde{a}^2\right]^2-\Tilde{a}^2\Delta\sin^2\theta
    \label{eq:SDA}
\end{equation}
with $\Tilde{a}=a\sqrt{1+l}$. Here, $l$ is called the measure of spotaneous Lorentz symmetry breaking, and $b$ denotes the charge parameter. In the context of Einstein-Bumblebee gravity, $b$ can be regarded as the vacuum expectation value of the Bumblebee field, and $l\propto \varrho b^2/a$, where $\varrho$ is the coupling constant between the gravitational field and the Bumblebee field \cite{Jha:2020pvk,Ding:2019mal}. The event horizon and ergosphere are located at
\begin{equation}
    \begin{split}
        r_{\rm EH}=& 1-\frac{b}{2}+\frac{\sqrt{(b-2)^2-4\Tilde{a}^2}}{2} \\
        r_{\rm ergo}=& 1-\frac{b}{2}+\frac{\sqrt{(b-2)^2-4\Tilde{a}^2\cos^2\theta}}{2}
    \end{split}
    \label{eq:rEH}
\end{equation}
On the equatorial plane, where $\theta=\pi/2$, we simply have $r_{\rm ergo,\frac{\pi}{2}}=2-b$. It is clear that the event horizon exists if and only if
\begin{equation}
    2-b\geq 2|a|\sqrt{1+l}
    \label{eq:bh_condition}
\end{equation}
The irreducible mass can be obtained by calculating the surface area of event horizon. The result is\footnote{Recall that we adopt the unit system with $M=1$}:
\begin{equation}
    M_{\rm irr}=\frac{1}{2}\sqrt{r_{\rm EH}\left(r_{\rm EH}+b\right)+\Tilde{a}^2}=\frac{1}{2}\sqrt{2-b+\mathfrak{r}}
    \label{eq:Mirr}
\end{equation}
Here we denote:
\begin{equation}
    \mathfrak{r}=\sqrt{(b-2)^2-4a^2(1+l)}\equiv 2r_{\rm EH}-r_{\rm ergo,\frac{\pi}{2}}
    \label{eq:frak_r}
\end{equation}
for simplicity. The condition for the existence of the event horizon, given in Eq~\eqref{eq:bh_condition}, is equivalent to $\mathfrak{r}^2\geq 0$. One can verify that the result in Eq.~\eqref{eq:Mirr} reduces to the case of the Kerr black hole when $l$ and $b$ vanish simultaneously. It is also evident that $M_{\rm irr}<1$.  Therefore, for a Kerr–Sen–like black hole, a significant amount of energy remains extractable, as in the case of a Kerr black hole.

As a stationary, axisymmetric metric, Kerr–Sen–like metric can be expressed in the 3+1 formalism as shown in Eq.~\eqref{eq:line}. In general, we have \cite{Carleo:2022qlv}:
\begin{equation}
    \alpha=\sqrt{1-\frac{2r}{\Sigma}+\frac{4\Tilde{a}^2r^2\sin^2\theta}{\Sigma A}}~~~,~~~\beta^r=\beta^{\theta}=0~~~,~~~
    \beta^{\phi}=\frac{2\Tilde{a}r\sin\theta}{\alpha\sqrt{\Sigma A}}
    \label{eq:alpha_beta}
\end{equation}
and $h_i=\sqrt{g_{ii}}$. The rest frame of ZAMOs can be defined via the tetrad as:
\begin{equation}
    \hat{e}_{(t)}^{\mu}=\frac{1}{\alpha}\left(\partial_t^{\mu}+\omega^{\phi}\partial_{\phi}^{\mu}\right),~~
    \hat{e}_{(r)}^{\mu}=\frac{1}{h_r}\partial_r^{\mu},~~~~
    \hat{e}_{(\theta)}^{\mu}=\frac{1}{h_{\theta}}\partial_{\theta}^{\mu},~~
    \hat{e}_{(\phi)}^{\mu}=\frac{1}{h_{\phi}}\partial_{\phi}^{\mu}.
    \label{eq:ZAMOs}
\end{equation}
through which one could determine the rest frame of the fluid moving on the equatorial plane, based on Eq.~\eqref{eq:fluid}.

\subsection{Geodesics}
\label{sec:geodesic}

There are three conserved quantities along geodesics in the Kerr–Sen–like spacetime: energy $E$, angular momentum $L$ and Carter constant $\mathcal{Q}$. In this work, we focus on geodesics confined to the equatorial plane, for which $\mathcal{Q}=0$. Given $E$ and $L$, the general form of geodesics on the equatorial plane could be expressed as:
\begin{equation}
    \begin{split}
        \Sigma U^t=& \frac{A}{\Delta}E-\frac{2\Tilde{a}r}{\Delta}L \\
        \Sigma U^r=& \pm\sqrt{\mathfrak{R}} \\
        \Sigma U^{\theta}=& 0 \\
        \Sigma U^{\phi}=&\frac{2\Tilde{a}r}{\Delta}E+\frac{\Sigma-2r}{\Delta}L
    \end{split}
    \label{eq:geodesic}
\end{equation} 
where
\begin{equation}
    \mathfrak{R}=\frac{1}{1+l}\left\{\left[(\Delta+2r)E-\Tilde{a}L\right]^2-(L-\Tilde{a}E)^2\Delta-m^2\Delta\Sigma)\right\}x
    \label{eq:R1}
\end{equation}
Here, we have $m=1$ for timelike geodesics, while $m=0$ for null geodesics. In alignment with the discussions in Ref.~\cite{Work1}, two types of timelike geodesics on the equatorial plane are considered. The first one is the circular orbit, which requires $\mathfrak{R}=\partial_{r}\mathfrak{R}=0$. The conserved quantities $E$ and $L$ on the circular orbits, expressed as functions of the radial coordinate $r$, satisfy:
\begin{equation}
    \begin{split}
        E_{\rm K}=&\left[\frac{4\sqrt{2}\Tilde{a}^3\Pi_2^{1/2}+4\sqrt{2}\Tilde{a}r\left(\Pi_1-2\right)\Pi_2^{1/2}-\left(\Pi_1-2\right)^2\Pi_2\Xi_1+2\Tilde{a}^2\Xi_2}{\Pi_1\left(8\Tilde{a}^2\Pi_2-\Xi_1^2\right)}\right]^{1/2}  \\
        L_{\rm K}=&\frac{\sqrt{2}\Delta\Pi_1\Pi_2^{1/2}-2\Tilde{a}\Xi_3-2\Tilde{a}^3}{\left(\Pi_1-2\right)^2\Pi_2-2\Tilde{a}^2}\times E_{\rm K}
    \end{split}
    \label{eq:EL_Kep}
\end{equation}
where
\begin{equation}
    \begin{split}
        \Pi_1&=r+b~~~,~~~\Pi_2=2r+b \\
        \Xi_1&=2r^2+3(b-2)r+b(b-2) \\
        \Xi_2&=6r^2+(9b-10)r+3b(b-2)~~,~~\Xi_3=3r^2+4(b-1)r+b(b-2)
    \end{split}
\end{equation}
Similar to the case of Kerr spacetime, the circular orbits exist from infinity down to the photon sphere $r_{\rm ph}$, where $E_{\rm K}\rightarrow \infty$ and $L_{\rm K}\rightarrow -\infty$. The radius of the photon sphere can be determined by solving $\mathfrak{R}=\partial_{r}\mathfrak{R}=0$ with $m=0$. The circular orbits are unstable at radii smaller than that of innermost stable circular orbit (ISCO) $r_{\rm ms}$, obtained by solving $\mathfrak{R}=\partial_r\mathfrak{R}=\partial_r^2\mathfrak{R}=0$. Both $r_{\rm ph}$ and $r_{\rm ms}$ can be numerically determined after $a$, $l$, and $b$ are specified. We will show in Sect.~\ref{sec:prop} that $r_{\rm EH}<r_{\rm ph}<r_{\rm ms}$ always holds whenever Eq.~\eqref{eq:bh_condition} is satisfied.

\section{Results}
\label{sec:res}

We will present our results in this section. We begin by briefly discussing the properties of the Kerr–Sen–like black hole. We will exhibit the regions where the event horizon exists in the $l–b$ plane, and how the radii of the photon sphere and the ISCO vary with $l$ and $b$. Next, we will show the allowed regions for energy extraction (regions with $\eta>1$) on different parameter planes. Furthermore, the distribution of covering factor, the value that quantifies the capability of an accretion system on energy extraction via magnetic reconnection, on $l–b$ plane will be plotted. By analyzing the variations of the allowed regions and the distributions of the covering factor, we aim to understand the influences of $l$ and $b$ on energy extraction.

\subsection{Properties of Kerr–Sen–like black hole}
\label{sec:prop}

\begin{figure}
    \centering
    \includegraphics[width=0.48\textwidth]{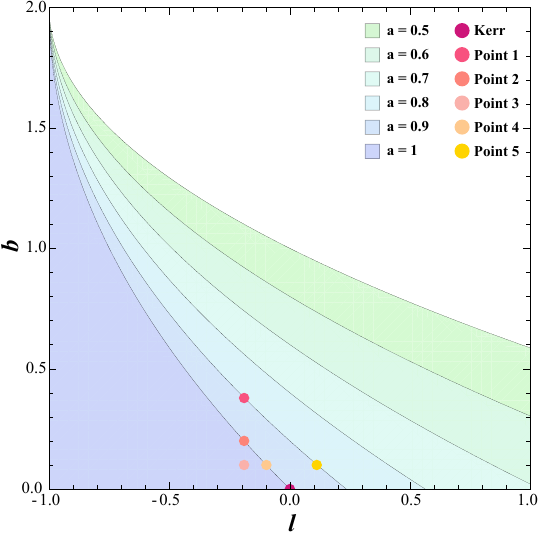}
\caption{The regions in which the event horizon exists are colored from green to blue for various values of $a$ in the $l–b$ parameter plane. The solid circles denote the sets of $(l,b)$ that are chosen as example sets}
    \label{fig:lb}
\end{figure}

\begin{table}[htbp]
    \centering
    \begin{tabular}{c|cccccc}
        \hline
        ~~ & Kerr & Point 1 & Point 2 & Point 3 & Point 4 & Point 5  \\
        \hline
        $l$ & 0 & -0.19 & -0.19 & -0.19 & -0.98 & 0.11  \\
        $b$ & 0 & 0.38 & 0.20 & 0.10 & 0.10 & 0.10  \\
        \hline
    \end{tabular}
    \caption{Values of $(l,b)$ for the example sets.}
    \label{tab:example}
\end{table}

From the definitions of parameters and form of metric, we directly obtain $0<b<2$ and $l>-1$. In particular, taking M87* to be a Kerr–Sen–like black hole, an upper bound on $l$ was estimated to be $0.63$ \cite{Jha:2021eww,Wang:2021gtd}. From an analytical perspective, in this work, we restrict our analysis to the parameter ranges of $b\in [0,2]$ and $l\in (-1,1]$.

As we mentioned in Sect.~\ref{sec:metric}, Eq.~\eqref{eq:bh_condition} should be satisfied for the existence of an event horizon, ensuring that no naked singularity arises. In Fig.~\ref{fig:lb}, in the $l–b$ parameter plane, the regions where Eq.~\eqref{eq:bh_condition} is satisfied for corresponding black hole spins are colored from green to blue. These uncolored regions correspond to the cosmic censorship hypothesis is violated, and should be excluded in the $l–b$ plane in the following discussions. Generally, the region for higher black hole spin is entirely contained within that for a lower black hole spin. It means the allowed range of $(l,b)$ for the existence of event horizon are more restrictive as the central black hole spins faster. For the following discussions, we pick 6 sets of $(l,b)$ to be the example sets, denoted by solid circles in Fig.~\ref{fig:lb}. The case of a pure Kerr black hole, where $(l,b)=(0,0)$, is included. The values of selected $(l,b)$ for the example cases are provided in Table~\ref{tab:example}. From Point 1 to 3, $l$ remains constant while $b$ decreases. From Point 3 to 5, $b$ remains constant while $l$ increases. We will investigate the influence of $l$ and $b$ on energy extraction by comparing the results across the example sets.

\begin{figure}
    \centering
    \includegraphics[width=0.48\textwidth]{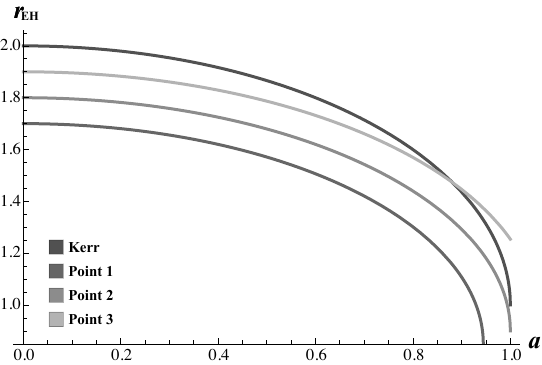}
    \hspace{2mm}
    \includegraphics[width=0.48\textwidth]{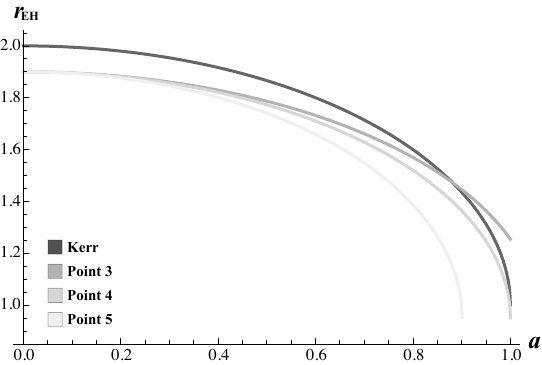}
    \includegraphics[width=0.48\textwidth]{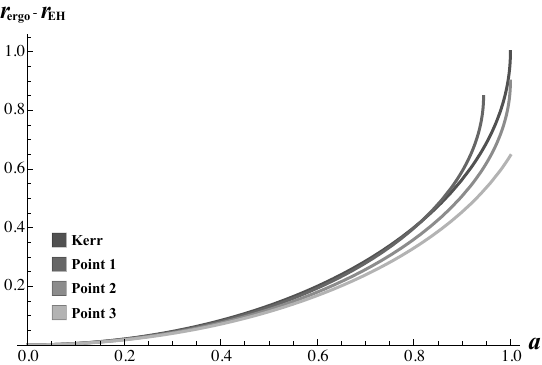}
    \hspace{1mm}
    \includegraphics[width=0.48\textwidth]{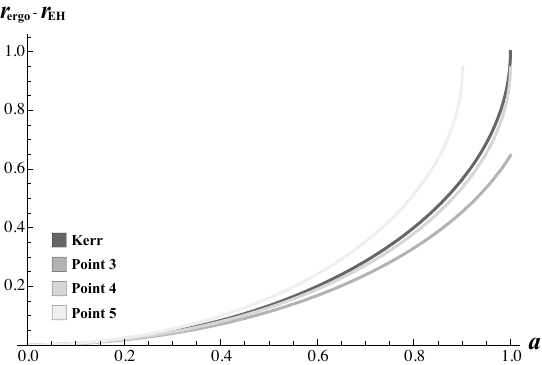}
    \includegraphics[width=0.48\textwidth]{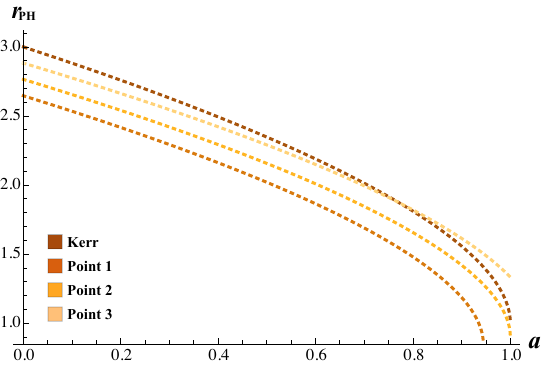}
    \hspace{2mm}
    \includegraphics[width=0.48\textwidth]{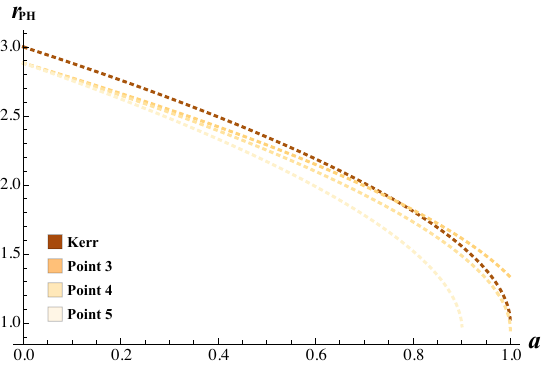}
    \includegraphics[width=0.48\textwidth]{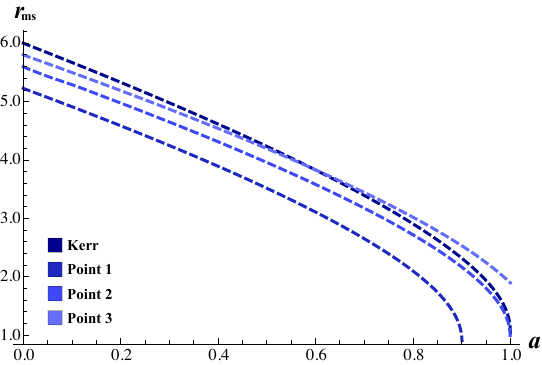}
    \hspace{2mm}
    \includegraphics[width=0.48\textwidth]{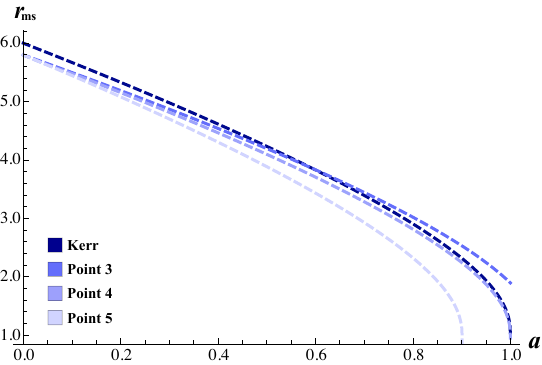}
    \caption{Radii of the event horizon (first row), photon sphere (third row), ISCO (fourth row), and the radial range of the ergoregion (second row) as functions of black hole spin for the cases of example sets.}
    \label{fig:rs}
\end{figure}

One can analyze Eq.~\eqref{eq:rEH} to study the influence of $l$ and $b$ on the position of the event horizon and the radial range of the ergoregion. On the equatorial plane, where $\theta=\pi/2$ so that $r_{\rm{ergo},\frac{\pi}{2}}=2-b$, we have:
\begin{equation}
    \begin{aligned}
        \delta r_{\rm EH} &= -\frac{a^2}{\mathfrak{r}}\delta l-\frac{r_{\rm EH}}{\mathfrak{r}}\delta b \\
        \delta \left(r_{\rm{ergo},\frac{\pi}{2}}-r_{\rm EH}\right) &= \frac{a^2}{\mathfrak{r}}\delta l+\frac{\left(r_{\rm{ergo},\frac{\pi}{2}}-r_{\rm EH}\right)}{\mathfrak{r}}\delta b
    \end{aligned}
    \label{eq:delta-r}
\end{equation}
We can see from Eq.~\eqref{eq:delta-r} that $r_{\rm EH}$ is negatively correlated with both $l$ and $b$, since $\mathfrak{r}\geq 0$ whenever the event horizon exists. This implies that the event horizon shifts inward as either $l$ or $b$ increases. Meanwhile, the second equation in Eq.~\eqref{eq:delta-r} indicates that increasing $l$ or $b$ broadens the radial range of the ergoregion on the equatorial plane. A larger radial range of the ergoregion caused by a bigger $l$ arises purely from the shrinkage of event horizon. In contrast, when $b$ increases, the radial range of the ergoregion broadens because the inward movement of the ergosphere is slower than that of the event horizon.

From the panels in the first row of Fig.~\ref{fig:rs}, one can clearly see the influence of $l$ and $b$ on $r_{\rm EH}$ as a function of $a$. By comparing the $r_{\rm EH}$ of Points 3, 4 and 5, we find that the effect of $l$ on $r_{\rm EH}$ becomes more and more considerable when $a$ increases. Similarly, the comparison among Points 1, 2 and 3 shows that the effect of $b$ on $r_{\rm EH}$ becomes slightly less significant as $a$ increases. The variations of $r_{\rm ph}$ and $r_{\rm ms}$ with respect to $l$ and $b$ follow similar trends to those of $r_{\rm EH}$, as shown in the third and fourth rows of Fig.~\ref{fig:rs}. In the second row of Fig.~\ref{fig:rs}, we plot $r_{\rm{ergo},\frac{\pi}{2}}-r_{\rm EH}$ as a functions of $a$. Obviously, we can see that the radial range of ergoregion grows with increasing $l$ or $b$. Additionally, the influence of $l$ or $b$ on $r_{\rm{ergo},\frac{\pi}{2}}-r_{\rm EH}$ is considerable only when $a$ is sufficiently large. All the results shown in Fig.~\ref{fig:rs} are consistent with what is implied by Eq.~\eqref{eq:delta-r}.

\begin{figure}
    \centering
    \hspace{-2mm}
    \includegraphics[width=0.48\textwidth]{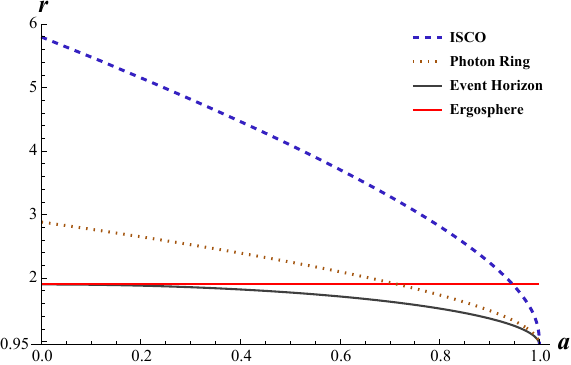}
    \hspace{2mm}
    \includegraphics[width=0.48\textwidth]{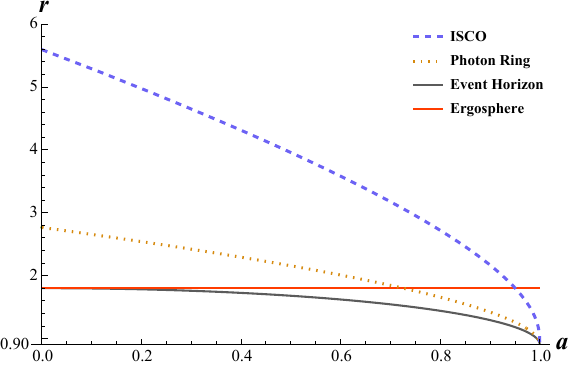}
    \caption{Radii of event horizon (black solid lines), ergosphere (red solid lines), photon sphere (gold dotted lines) and ISCO (azure dashed lines) as functions of black hole spin in the cases of Point 1 (left panel) and 2 (right panel), respectively.}
    \label{fig:rp}
\end{figure}

To examine the relationships among the radii of the event horizon, photon sphere, and ISCO, we plot them as functions of black hole spin for Points 1 and 2 separately in Fig.~\ref{fig:rp}. It is evident that, for whatever values of $l$ and $b$ (with existence of event horizon), the inequality $r_{\rm EH}< r_{\rm ergo}$ strictly holds in Kerr–Sen–like spacetime. Meanwhile, as $a\rightarrow 0$, $r_{\rm EH}$ approaches $r_{\rm ergo}$, causing the ergoregion to gradually disappear. Additionally, the relation $r_{\rm EH}<r_{\rm ph}<r_{\rm ms}$ always holds. As $a\rightarrow 1$, we have $r_{\rm EH}=r_{\rm ph}=r_{\rm ms}$. The relationships among these three radii in Kerr–Sen–like spacetime are similar to those in pure Kerr spacetime.

To investigate the effects of $l$ and $b$ on the irreducible mass, one should revisit Eq.~\eqref{eq:Mirr}, which gives:
\begin{equation}
    \sqrt{2}\times\delta M_{\rm irr}=-\frac{a^2}{\mathfrak{r}\sqrt{r_{\rm EH}}}\delta l-\frac{\sqrt{r_{\rm EH}}}{\mathfrak{r}}\delta b
\end{equation}
Hence, the irreducible mass of a Kerr–Sen–like black hole decreases with increasing $l$ or $b$. In other words, more energy could be extracted from the rotating black hole if there was larger rate of Lorentz symmetry breaking or Bumblebee charge in the spacetime. For a fixed black hole spin, a constant irreducible mass corresponds to a line on the $l–b$ plane of the form:
\begin{equation}
    a^2l+2M_{\rm irr}^2b=const=-4M_{\rm irr}^4+4M_{\rm irr}^2-a^2
    \label{eq:M_contour}
\end{equation}
We temporarily call a line satisfying Eq.~\eqref{eq:M_contour} in $l–b$ plane a contour line of irreducible mass for a fixed black hole spin. It is not hard to figure out that a contour line of irreducible mass is tangent to the boundary defined by Eq.~\eqref{eq:bh_condition} at the point:
\begin{equation}
    l=\frac{M_{\rm irr}^2}{a^2}-1~~~~,~~~~b=2-4M_{\rm irr}^2
    \label{eq:tangency}
\end{equation}
One should notice that the variations of irreducible mass with $l$ and $b$ are not well defined on this point, since $\mathfrak{r}\equiv 0$. We will discuss this point of tangency further in Sect.~\ref{sec:covering} after presenting the variations of covering factor along the contour lines of irreducible mass.

\subsection{Allowed regions for energy extraction}
\label{sec:allow}

\begin{figure}
    \centering
    \hspace{-2mm}
    \includegraphics[width=0.48\textwidth]{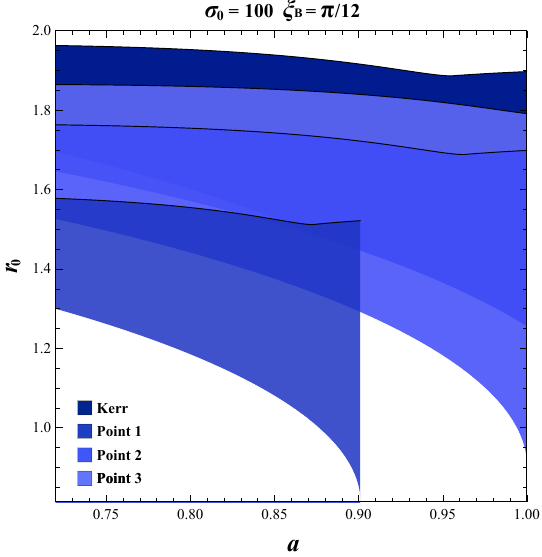}
    \hspace{2mm}
    \includegraphics[width=0.48\textwidth]{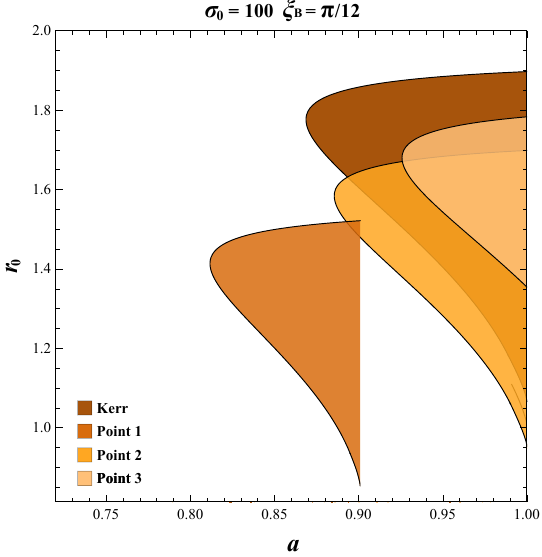}
    \hspace{-2mm}
    \includegraphics[width=0.48\textwidth]{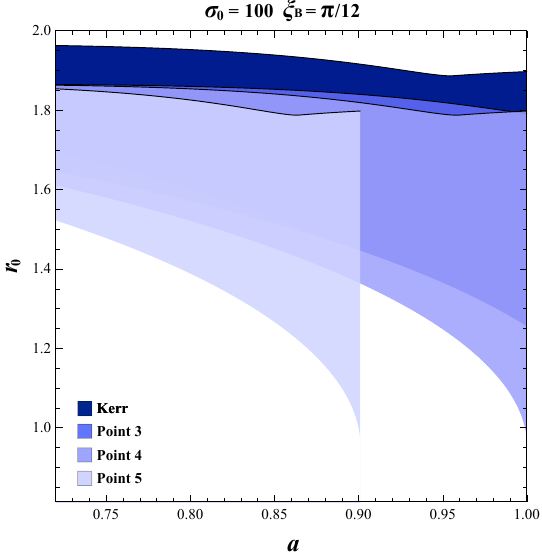}
    \hspace{2mm}
    \includegraphics[width=0.48\textwidth]{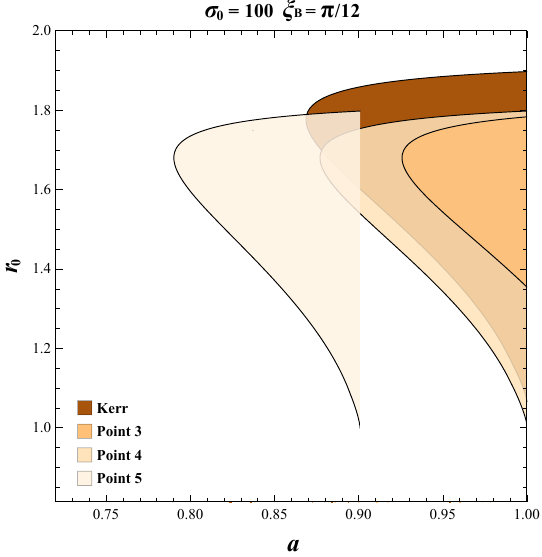}
    \caption{The $a–r_0$ parameter planes for the combined (left panels) and circular (right panels) streamlines. The allowed regions for energy extraction in the cases of example sets with $\sigma_0=100$ and $\xi_B=\pi/12$ are colored in azure (combined streamlines) and gold (circular streamlines). }
    \label{fig:a-r}
\end{figure}

We plot the allowed region for energy extraction via magnetic reconnection on $a–r_0$ planes for the example sets with $\sigma_0=100$ and $\xi_{\rm B}=\pi/12$ in Fig.~\ref{fig:a-r}. The allowed regions in the cases of combined and circular streamlines are (henceforth) colored in azure and gold, respectively. The allowed regions for Points 1, 2 and 3 are plotted on the upper panels, among which $l$ keeps constant but $b$ varies. While the allowed regions for Points 3, 4 and 5 are plotted on the bottom panels, among which $b$ remains constant but $l$ varies. The case of Kerr black hole is plotted on each panel as a comparison. Note that for Point 1 and 5, the maximally allowed value of $a$ is 0.9. To clearly display the allowed regions alongside the ergoregions, photon spheres, and ISCOs, we specifically plot the cases of Points 2 and 5 in the top and bottom panels of Fig.~\ref{fig:a-r-p}, respectively (see Appendix~\ref{sec:p25}). In this work, we do not discuss how the streamline of the bulk plasma and the orientation angle affect the allowed region, as these aspects have already been studied in detail in Ref.~\cite{Work1,Work2}

In Fig.~\ref{fig:a-r}, a comparison among Points 1, 2, and 3 on the upper panels reveals that the allowed region shifts downward in the $a–r_0$ plane as the charge parameter increases. This behavior is actually the consequence of the inward movements of the ergosphere and the event horizon. Furthermore, for a fixed black hole spin, the radial range of allowed region is larger when the charge parameter increases. On the other hand, a comparison among Points 3, 4 and 5, exhibited in the bottom panels of Fig.~\ref{fig:a-r}, clearly shows that increasing the rate of Lorentz symmetry breaking causes the allowed region in $a–r_0$ to shift markedly leftward. Actually, the allowed regions for Point 3, 4 and 5 are essentially the same. The three allowed regions overlap completely if they were plotted on the $\Tilde{a}–r_0$ parameter planes. It is reasonable, since the quantity $\Tilde{a}=a\sqrt{1+l}$ in Kerr–Sen–like metric plays the role of the effective black hole spin, as first discussed in Ref.~\cite{Khodadi:2022dff}.

\begin{figure}
    \centering
    \hspace{-2mm}
    \includegraphics[width=0.48\textwidth]{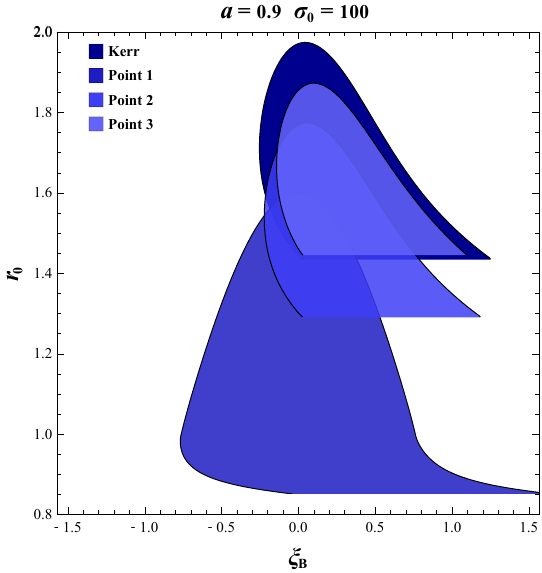}
    \hspace{2mm}
    \includegraphics[width=0.48\textwidth]{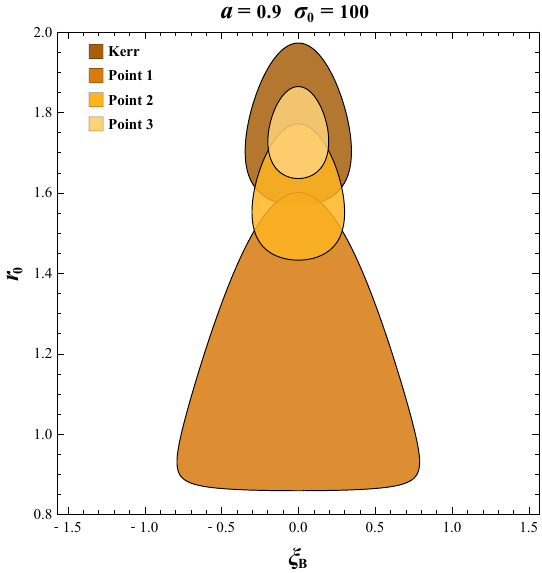}
    \\
    ~~~~~~~~~~~~
    \\
    \hspace{-2mm}
    \includegraphics[width=0.48\textwidth]{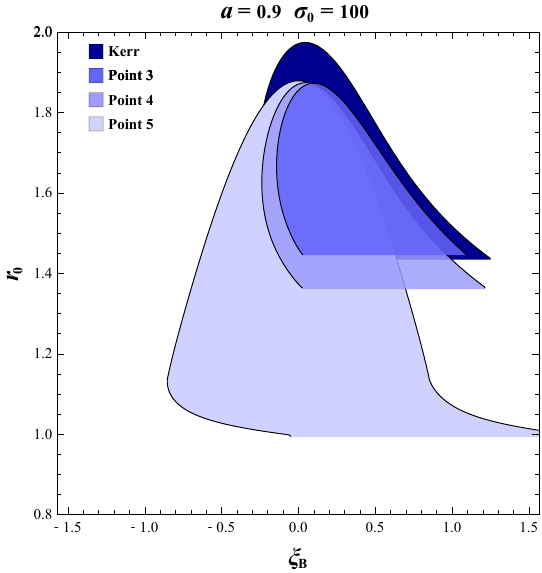}
    \hspace{2mm}
    \includegraphics[width=0.48\textwidth]{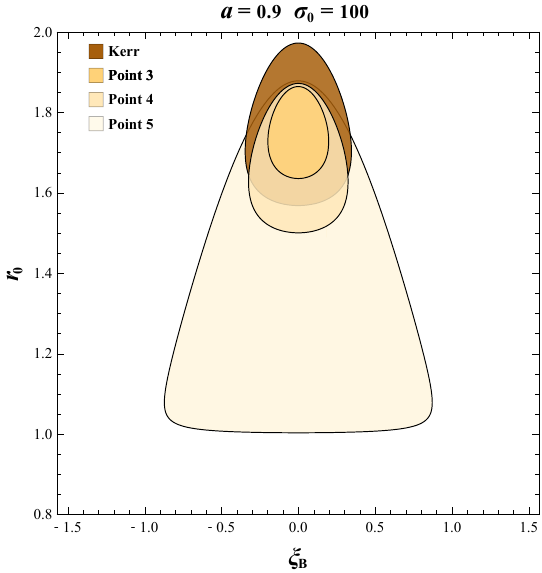}
    \caption{The $\xi_{\rm B}–r_0$ parameter planes for the combined (left panels) and circular (right panels) streamlines. The allowed regions for energy extraction in the cases of example sets with $\sigma_0=100$ and $a=0.9$ are colored in azure (combined streamlines) and gold (circular streamlines). }
    \label{fig:xi-r-a9}
\end{figure}

\begin{figure}
    \centering
    \hspace{-2mm}
    \includegraphics[width=0.48\textwidth]{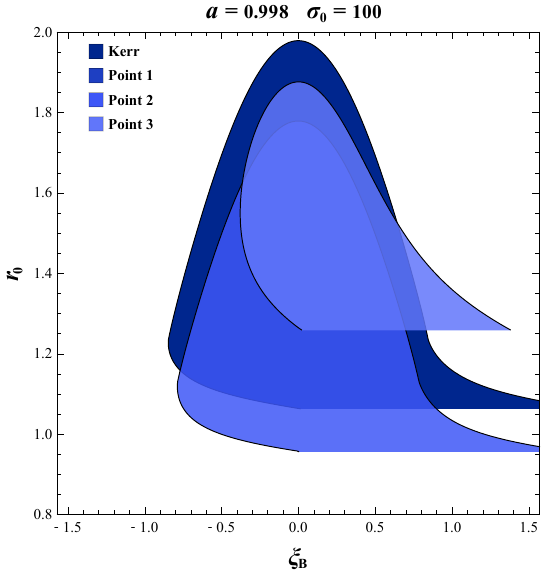}
    \hspace{2mm}
    \includegraphics[width=0.48\textwidth]{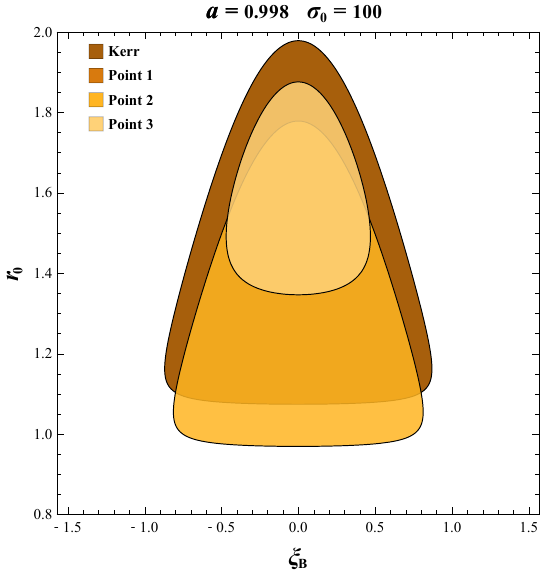}
    \\
    ~~~~~~~~~~~~
    \\
    \hspace{-2mm}
    \includegraphics[width=0.48\textwidth]{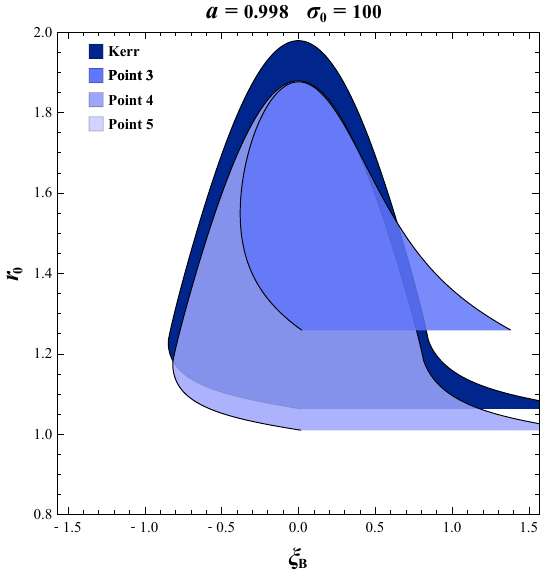}
    \hspace{2mm}
    \includegraphics[width=0.48\textwidth]{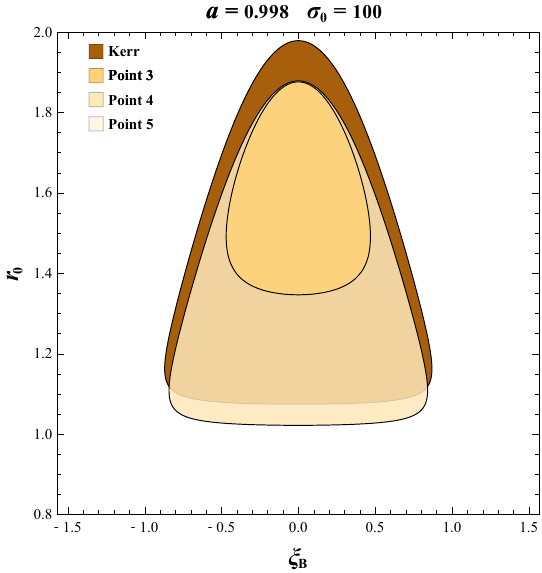}
    \caption{The $\xi_{\rm B}–r_0$ parameter planes for the combined (left panels) and circular (right panels) streamlines. The allowed regions for energy extraction in the cases of example sets with $\sigma_0=100$ and $a=0.998$ are colored in azure (combined streamlines) and gold (circular streamlines). }
    \label{fig:xi-r-a998}
\end{figure}

From an astrophysical perspective, people may be more interested in the scenario where the black hole spin is fixed. To this end, we plot the allowed regions for energy extraction in the $\xi_{\rm B}–r_0$ planes for example sets in Fig.~\ref{fig:xi-r-a9}, with $a=0.9$, $\sigma_0=100$, combined and circular streamlines, respectively. The corresponding plots for $a=0.998$ are shown in Fig.~\ref{fig:xi-r-a998}. Since $a=0.998$ is not allowed in the cases of Points 1 and 5, only four allowed regions appear in each panel of Fig.~\ref{fig:xi-r-a998}. The allowed regions in the cases of Points 2 and 5 with $a=0.9$ are specifically shown in Fig.~\ref{fig:xi-r-p}, where the ergoregions, photon spheres, ISCOs and the best orientation angle are displayed together (see Appendix~\ref{sec:p25}). 

As indicated by Fig.~\ref{fig:xi-r-a9} and Fig.~\ref{fig:xi-r-a998}, a larger charge parameter or a higher rate of Lorentz symmetry breaking results in a more spacious allowed region for energy extraction in the $\xi_{\rm B}-r_0$ plane. When $b$ increases, as shown in upper panels of Fig.~\ref{fig:xi-r-a9} and Fig.~\ref{fig:xi-r-a998}, both the upper and lower radial bounds shift inward. The upper boundary shifts inward as a result of the ergosphere contracting with increasing $b$. The lower radial bound moves inward with the inward movement of the photon sphere for the circularly flowing bulk plasma, and with that of the event horizon for the plunging bulk plasma. Hence, when the spacetime contains a larger Bumblebee charge, it is expected that energy extraction via magnetic reconnection would succeed at locations closer to the central black hole. When $l$ increases, as shown in the bottom panels of Fig.~\ref{fig:xi-r-a9} and Fig.~\ref{fig:xi-r-a998}, the lower radial bound of the allowed region shifts inward, while the upper radial bound remains nearly constant or shifts slightly outward due to the broadening of the allowed region. This behavior arises because the radius of the ergosphere on the equatorial plane is independent of $l$. Hence, when the rate of Lorentz symmetry breaking is larger in the spacetime, we may predict that successful energy extraction would also occur in regions closer to the center, which, for smaller $l$, may lie inside the event horizon.

One may also notice that the shapes of the allowed regions for the plunging bulk plasma change as $l$ or $b$ increases. The allowed regions in the cases of Points 1 and 5 in the left panels of Fig.~\ref{fig:xi-r-a9} appear markedly different from the others. This is because, in these cases, $b$ or $l$ is large enough that the ISCO, the upper radial boundary of the plunging region, moves into the ergoregion. As a result,the dependence of energy extraction on the orientation angle is subsequently altered. Similarly, the case of Point 3 in the left panels of Fig.~\ref{fig:xi-r-a998} appear different from others because the ISCO lies outside the ergoregion.

\subsection{Covering factor}
\label{sec:covering}

Although the allowed region for energy extraction broadens as $b$ or $l$ increases, it is incorrect to expect that energy extraction via magnetic reconnection becomes more likely to succeed, since the radial range of the ergoregion also broadens, allowing magnetic reconnection to occur randomly within a wider radial range. The question of whether energy extraction via magnetic reconnection would be more likely to succeed, with a larger charge parameter or a greater rate of Lorentz symmetry breaking, should be addressed properly by calculating the covering factor:
\begin{equation}
    \chi=\frac{S_{\rm EE}}{\pi\left(r_{\rm ergo,\pi/2}-r_{\rm EH}\right)}
    \label{eq:chi}
\end{equation}
where $S_{\rm EE}$ is the area of the allowed region for energy extraction in the $\xi_{\rm B}–r_0$ plane. The covering factor quantifies the capability of an accretion system to extract energy from its central black hole via magnetic reconnection. Assuming that magnetic reconnection occurs with equal probability at any radius within the ergosphere and along any orientation angle, the covering factor represents the probability of successful energy extraction from the central black hole when a magnetic reconnection process occurs in the ergoregion. We do not directly refer to this value as “the probability”, because the real-world situations are much more complicated. Nevertheless, the covering factor remains a useful measure in exploratory studies.

\begin{figure}
    \centering
    \hspace{-2mm}
    \includegraphics[width=0.48\textwidth]{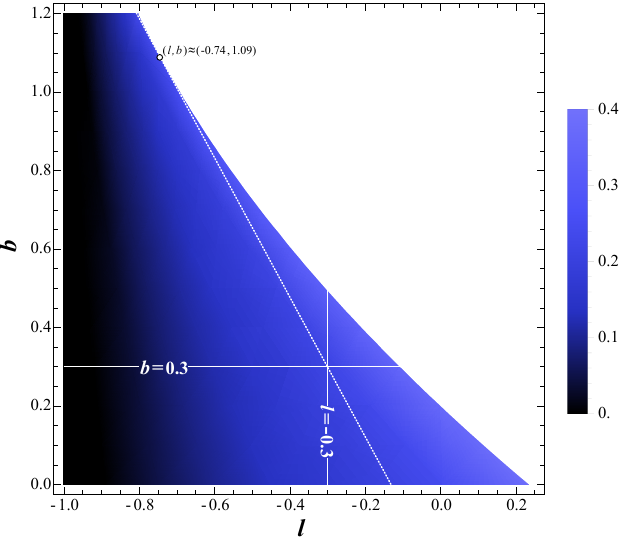}
    \hspace{2mm}
    \includegraphics[width=0.48\textwidth]{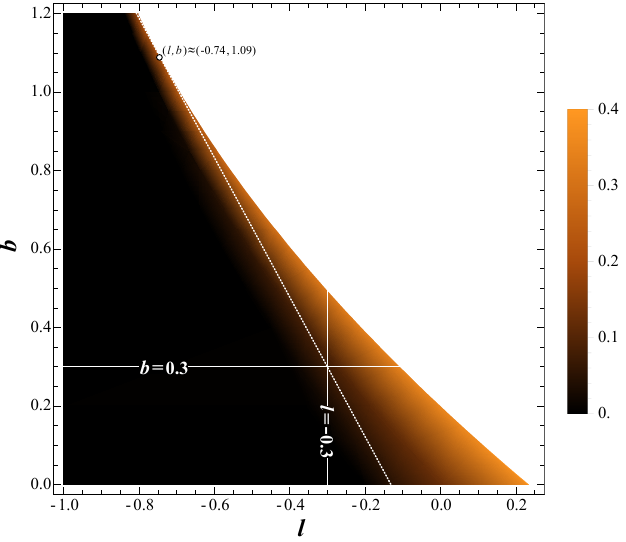}
    ~~~~~~~~~~~~~~~\\
    \hspace{-2mm}
    \includegraphics[width=0.45\textwidth]{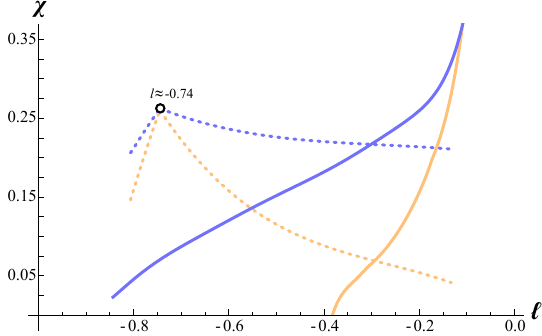}
    \hspace{6mm}
    \includegraphics[width=0.45\textwidth]{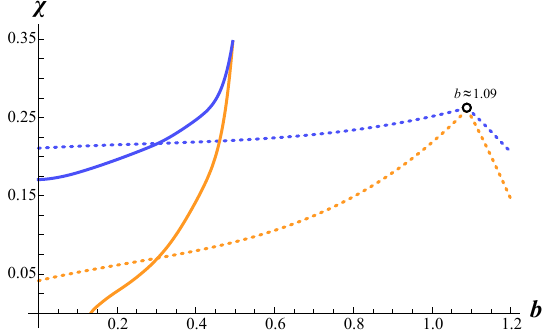}
    \caption{The distributions of the covering factor on the $l$–$b$ plane with $a=0.9$ are plotted in the top panels, for the cases of the combined (top left panel) and circular (top right panel) streamlines. The white dotted lines represent the contour lines of irreducible mass. The bottom left panel shows the variations of the covering factor with $l$ for $a=0.9$, along the line of $b=0.3$ (solid lines) and along the contour line of irreducible mass (dotted lines). Similarly, the bottom right panel illustrates the variations of the covering factor with $b$ along the line of $l=-0.3$ (solid lines) and along the contour line of irreducible mass (dotted lines).}
    \label{fig:chi-lb-a9}
\end{figure}

\begin{figure}
    \centering
    \hspace{-2mm}
    \includegraphics[width=0.48\textwidth]{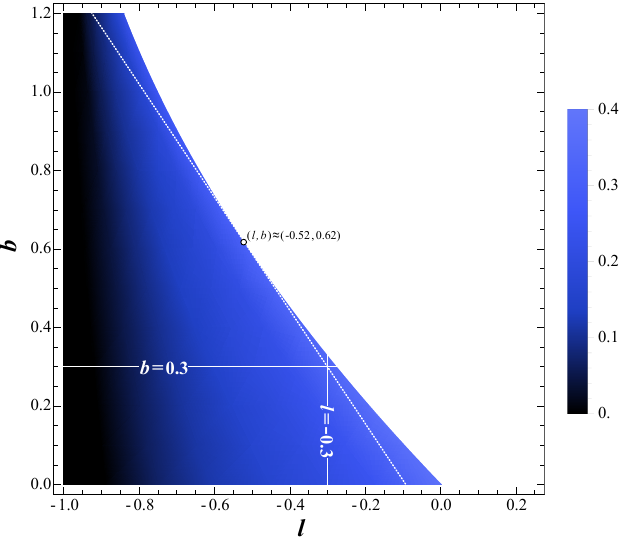}
    \hspace{2mm}
    \includegraphics[width=0.48\textwidth]{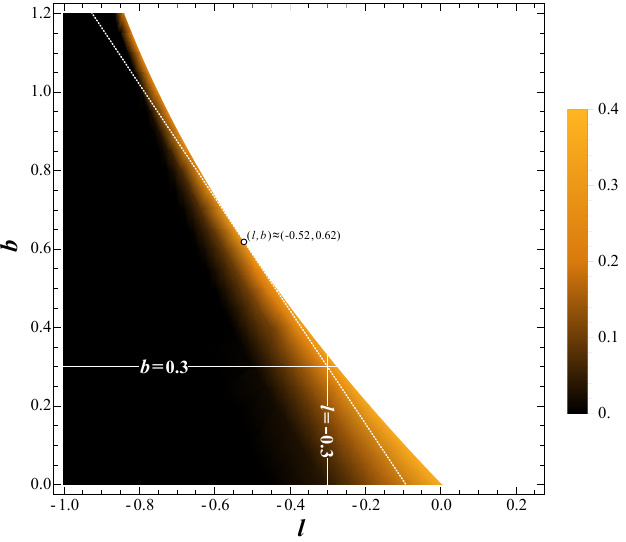}
    ~~~~~~~~~~~~~~~\\
    \hspace{-2mm}
    \includegraphics[width=0.45\textwidth]{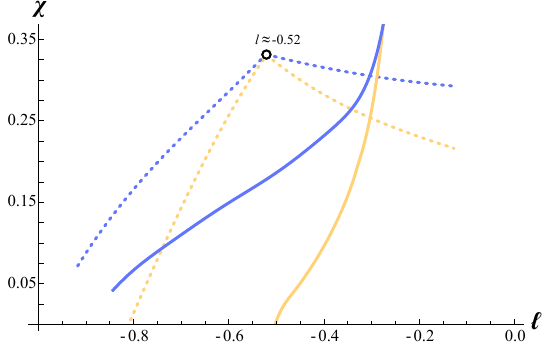}
    \hspace{6mm}
    \includegraphics[width=0.45\textwidth]{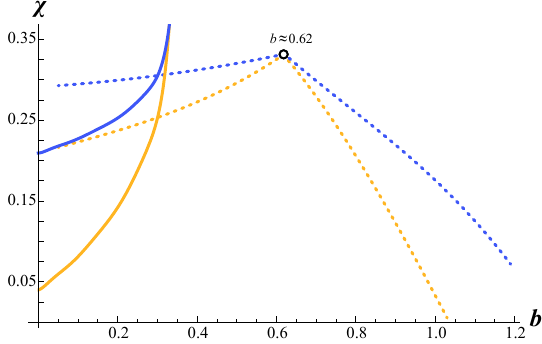}
    \caption{The distributions of the covering factor on the $l$–$b$ plane with $a=0.998$ are plotted in the top panels, for the cases of the combined (top left panel) and circular (top right panel) streamlines. The white dotted lines represent the contour lines of irreducible mass. The bottom left panel shows the variations of the covering factor with $l$ for $a=0.998$, along the line of $b=0.3$ (solid lines) and along the contour line of irreducible mass (dotted lines). Similarly, the bottom right panel illustrates the variations of the covering factor with $b$ along the line of $l=-0.3$ (solid lines) and along the contour line of irreducible mass (dotted lines).}
    \label{fig:chi-lb-a998}
\end{figure}

The distributions of the covering factor on the $l–b$ plane for $a=0.9$, $\sigma_0=100$ are plotted in the top panels of Fig.~\ref{fig:chi-lb-a9}, for the cases of the combined and circular streamlines, respectively. Meanwhile, we plot the variations of the covering factor with $l$ along the lines of $b=0.3$ (solid) and the contour lines of irreducible mass (dotted) in the bottom left panel of Fig.~\ref{fig:chi-lb-a9}. The variations with $b$ along the line of $l=-0.3$, along with the contour lines of irreducible mass, are shown in the bottom right panel. Corresponding plots for $a=0.998$ are shown in Fig.~\ref{fig:chi-lb-a998}. A clear conclusion is that an accretion system with a faster-spinning central black hole has a larger covering factor, as seen by comparing Fig.~\ref{fig:chi-lb-a9} and Fig.~\ref{fig:chi-lb-a998}.

From Fig.~\ref{fig:chi-lb-a9} and Fig.~\ref{fig:chi-lb-a998}, it is evident that the covering factor resulting from the combined streamline (colored in azure) is generally higher than that from the circular streamline (colored in gold) when other conditions are the same. It follows directly that an accretion system with plunging bulk plasma is more capable of extracting energy from the central black hole via magnetic reconnection than one with circularly flowing bulk plasma, which is consistent with the results in Ref.~\cite{Work0,Work1,Work2}. Furthermore, a closer inspection of the bottom panels of Fig.~\ref{fig:chi-lb-a9} and Fig.~\ref{fig:chi-lb-a998} reveals that the covering factor increases with increasing $l$ or $b$. This result is reasonable, as the ergoregion moves closer to the center of the black hole and encompasses a broader radial range as  $l$ or $b$ increases, as discussed in Sec.~\ref{sec:prop}. Consequently, energy extraction via magnetic reconnection is generally more likely to occur when the spacetime contains a higher rate of Lorentz symmetry breaking or a larger Bumblebee charge. It also appears that the variation of the covering factor with $l$ is comparable to that with $b$, indicating that the rate of Lorentz symmetry breaking and the charge parameter have similar influences on energy extraction.

The variations of the covering factor along the contour lines of irreducible mass are represented by dotted lines in the bottom panels of Fig.~\ref{fig:chi-lb-a9} and Fig.~\ref{fig:chi-lb-a998}. The point discussed in Eq.~\eqref{eq:tangency}, where the contour line of irreducible mass is tangent to the boundary for the existence of the event horizon, is marked by a tiny hollow circle in each panel. As argued in Sec.~\ref{sec:prop}, since $\delta M_{\rm irr}$ is ill-defined, the derivative of any function along the contour line of irreducible mass will never be continuous at this point. Hence, as shown in the bottom panels of Fig.~\ref{fig:chi-lb-a9} and Fig.~\ref{fig:chi-lb-a998}, the variations of the covering factor along the contour lines of irreducible mass at this point become nonsmooth.

As we know, two black holes with the same irreducible mass contain identical amounts of extractable energy. An interesting question then arises as to whether the difficulty of extracting energy via magnetic reconnection from the central black holes, as reflected by the covering factor, is the same if the black holes possess identical extractable energy. The answer for pure Kerr black holes is straightforward. Because the black hole spin is the only determining factor, it is equally difficult to extract energy via magnetic reconnection from central black holes if their irreducible masses are the same, provided that all other conditions of the accretion flow are identical. Moreover, energy extraction becomes easier when the central black hole contains more extractable energy. However, the answer is not as evident when we consider Bumblebee gravity, in which both the Bumblebee charge and the rate of Lorentz symmetry breaking are included.

This question could be answered by checking the variation of the covering factor along the contour lines of irreducible mass. As shown in Fig.~\ref{fig:chi-lb-a9} and Fig.~\ref{fig:chi-lb-a998}, according to our results, although two accretion systems may contain central black holes with the same extractable energy, the difficulty of energy extraction from their central black holes via magnetic reconnection are not necessarily the same. What is more, the energy of a central black hole is more likely to be extracted via magnetic reconnection when the internal properties of the spacetime (including the central black hole spin, rate of Lorentz symmetry breaking, and charge parameter) are closer to the boundary given by the cosmic censorship hypothesis. In other words, when the extractable energy of the central black hole is determined, energy extraction via magnetic reconnection is more favorable when the spacetime carries certain $(l,b)$ values such that the cosmic censorship hypothesis is marginally not violated.

\section{Summary and discussions}
\label{sec:sum}

In this paper, we extend our discussions in Ref.~\cite{Work0,Work1,Work2}, about energy extraction from a rotating black hole via magnetic reconnection, to the case of Bumblebee gravity. The foundation of our work is the Comisso–Asenjo process introduced in Ref.~\cite{CA2021}. Both the plunging and circularly flowing bulk plasmas are considered for comparison. The main results are consistent with those obtained in our previous works. 

We revisited the fundamentals of Bumblebee gravity, which yield a Kerr–Sen–like spacetime, and we also examined the Comisso–Asenjo process. Through analyses and plots, we determined that an increase in either the rate of Lorentz symmetry breaking or Bumblebee charge leads to a contraction of the event horizon and an expansion of the radial range. As $b$ increases, the event horizon moves inward faster than the ergosphere. When $l$ increases, the event horizon moves inward without affecting the ergosphere. Meanwhile, it was shown that the allowed regions for energy extraction broaden and move inward as the rate of Lorentz symmetry breaking or Bumblebee charge increases.

Furthermore, we plotted the variations of the covering factor, which quantifies the capability of an accretion system to extract energy from its central black hole. Generally, a larger $l$ or $b$ in the spacetime results in a larger covering factor, indicating that energy extraction via magnetic reconnection is more likely to succeed when the spacetime contains a larger rate of Lorentz symmetry breaking or Bumblebee charge. The influences of $l$ and $b$ on covering factor are similar, at least in high-spin cases, according to our results. Moreover, we examined the variation of the covering factor when the irreducible mass of the central black hole is determined. According to our results, for two accretion systems with central black holes containing the same extractable energies, black hole spins, and bulk plasma, the probabilities of extracting energy via magnetic reconnection from their central black holes are not necessarily the same. A maximal value of the covering factor is reached when $2-b-2|a|\sqrt{1+l}\rightarrow 0$. In other words, energy extraction via magnetic reconnection would be most likely to succeed if the black hole spin, rate of Lorentz symmetry breaking, and Bumblebee charge approach certain values such that the cosmic censorship hypothesis is marginally not violated.

One after another, advanced theories beyond pure Einstein gravity were proposed, among which Bumblebee gravity is one of the most promising to help us explore the mysteries of the fundamental physics in the future. \cite{Kostelecky:2003fs,Bailey:2006fd,Mai:2024lgk,Ji:2024aeg}. Investigations were conducted to estimate the rate of Lorentz symmetry breaking and the vacuum expectation value of the Bumblebee field near certain black holes \cite{Zhu:2024qcm,Jha:2021eww,Wang:2021gtd,Islam:2024sph,Gu:2022grg}. We hope this work provides a new perspective, aiding in the search for evidence of the Bumblebee field and motivating further analyses of Kerr–Sen–like spacetimes.

Although this work is the last study of a serial (including Ref.~\cite{Work0,Work1,Work2} done previously), it should never be the last progress of the Comisso–Asenjo process. Many problems remain to be solved, and numerous aspects still need to be optimized. For example, as we mentioned, the covering factor was chosen only temporarily to quantify the capability of energy extraction. Better choices should be made in the future after carefully considering the more complicated astrophysical environment of the accretion flow, Ref.~\cite{Camilloni:2024tny} was striving for. 

Hitherto, analytical works about magnetic reconnection near black holes stagnate at the study about “one-off” process. However, as we know, reconnections could happen drastically and violently under the effects of tearing and kink mode \cite{tearing-mode,Ebrahimi:2016urr,Sironi:2025kgn,Tsung:2025eku}. We hope that one day we could be able to handle the serial occurrences of magnetic reconnection in astrophysical systems through methods other than numerical simulations. We are also interested in how different the energy extraction would be if the reconnection process were treated in a more realistic, non-isolated way. To be honest, the gap between theoretical work and observation remains significant, even with the assistance of numerical simulations. Strenuous efforts must be made, step by step, to build a bridge connecting the theory of magnetic reconnection to the real dynamical processes in astrophysical systems.

\section{Acknowledgement}

We appreciate valuable suggestions given by Prof.~Bin Chen (NBU) and Prof.~Lijing Shao (KIAA). Ye Shen would like to thank Prof.~Minyong Guo (BNU) for suggesting this topic. The work is partly supported by NSFC Grant No. 12275004.

\appendix

\section{Allowed regions for Point 2 and 5}
\label{sec:p25}

\begin{figure}
    \centering
    \hspace{-2mm}
    \includegraphics[width=0.48\textwidth]{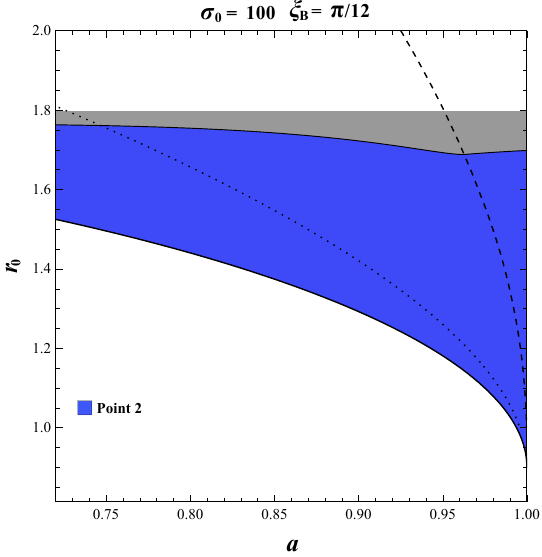}
    \hspace{2mm}
    \includegraphics[width=0.48\textwidth]{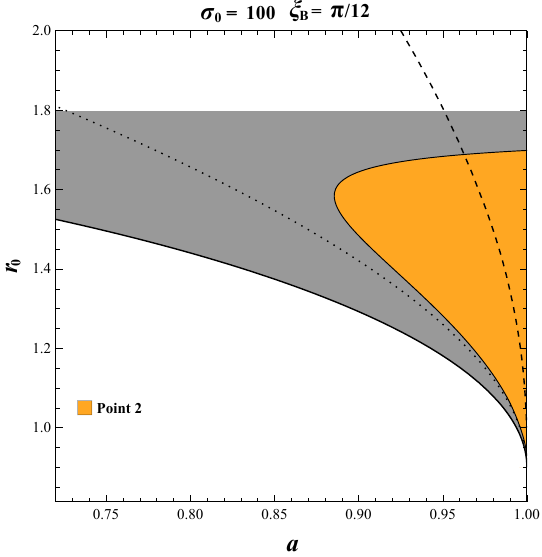}
    \\
    ~~~~~~~~~~~~
    \\
    \hspace{-2mm}
    \includegraphics[width=0.48\textwidth]{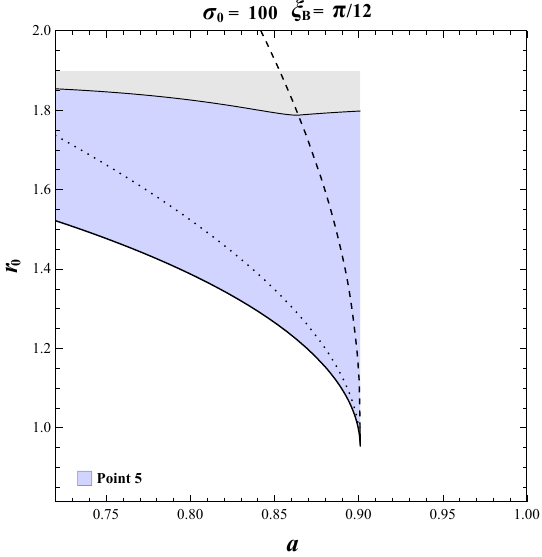}
    \hspace{2mm}
    \includegraphics[width=0.48\textwidth]{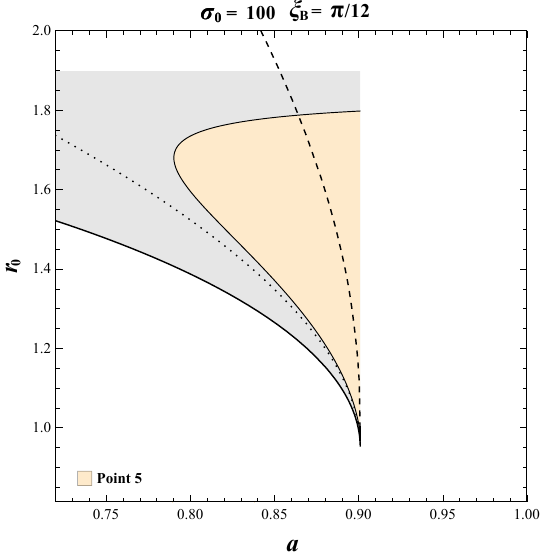}
    \caption{The $a–r_0$ parameter planes for the combined (left panel) and circular (right panel) streamlines, where the allowed regions with $\sigma_0=100$ and $\xi_B=\pi/12$, along with the ergoregions, are shown for the cases of Point 2 (upper panel) and Point 5 (lower panel). The solid, dotted, and dashed black lines represent the event horizon, photon sphere, and ISCO, respectively.}
    \label{fig:a-r-p}
\end{figure}

\begin{figure}
    \centering
    \hspace{-2mm}
    \includegraphics[width=0.48\textwidth]{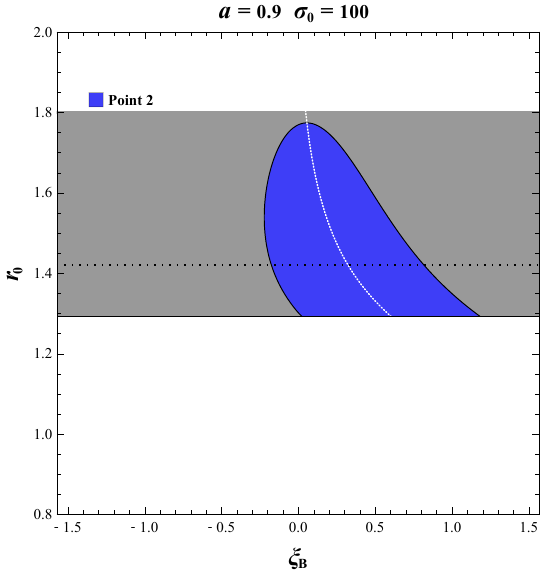}
    \hspace{2mm}
    \includegraphics[width=0.48\textwidth]{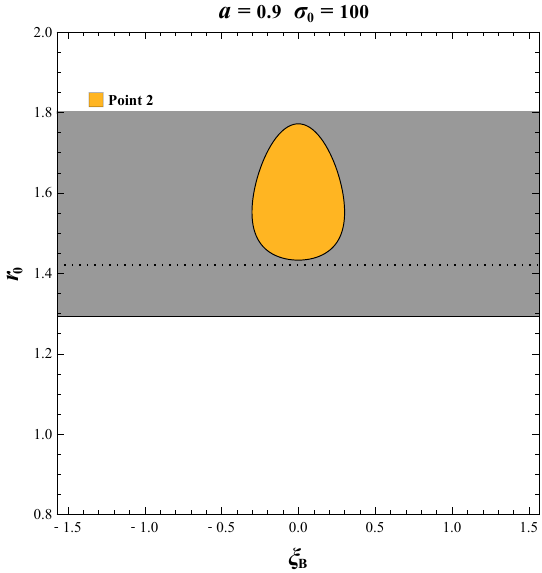}
    \\
    ~~~~~~~~~~~~
    \\
    \hspace{-2mm}
    \includegraphics[width=0.48\textwidth]{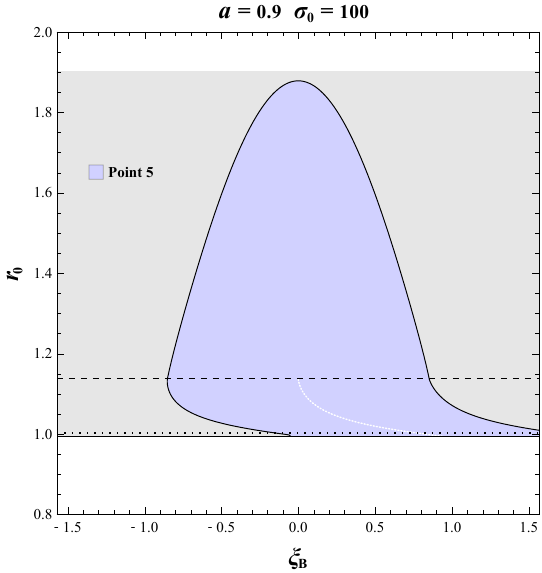}
    \hspace{2mm}
    \includegraphics[width=0.48\textwidth]{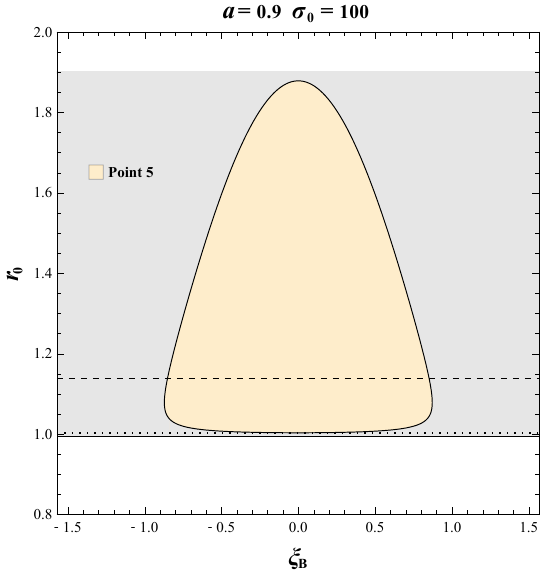}
    \caption{The $\xi_{\rm B}–r_0$ parameter planes for the combined (left panel) and circular (right panel) streamlines, where the allowed regions and ergoregions with $a=0.998$ are shown for the cases of Point 2 (upper panels) and Point 5 (lower panels). The solid, dotted, and dashed black lines represent the event horizon, photon sphere, and ISCO, respectively. The dotted white lines in the left panels represent the best orientation angle.}
    \label{fig:xi-r-p}
\end{figure}

It is not convenient to plot the ergoregions and all the characteristic radii for every case of example sets in the parameter planes shown previously. Hence, as a supplement, this section specifically presents the allowed regions, along with the ergoregion (colored in grey), the photon sphere (denoted by dotted lines), and the ISCO (also denoted by dotted lines), for the cases of Points 2 and 5. In Fig.~\ref{fig:a-r-p}, the allowed regions in the cases of Points 2 (upper panels) and 5 (lower panels) in the $a–r_0$ planes are shown. Meanwhile, the allowed regions in $\xi_{\rm B}–r_0$ planes are shown in Fig.~\ref{fig:xi-r-p}, where we also plot the best orientation angle (defined in Eq.~\eqref{eq:xi_m}) by white dotted lines in the plunging region on the left panels. With the help of these figures, one can more clearly observe the relationships among the allowed regions, ergoregions, all the characteristic radii, and even the best orientation angle.

In this section, we briefly summarize some properties of the allowed regions for energy extraction in the $a–r_0$ and $\xi_{\rm B}–r_0$ parameter planes, which have been discussed in greater detail in Ref.~\cite{Work1}  for the Kerr spacetime.

The allowed regions for energy extraction are generally more spacious when the bulk plasma plunges, as can be seen by comparing the left and right panels in Fig.~\ref{fig:a-r-p}. From the right panels of Fig.~\ref{fig:a-r-p}, where the bulk plasma flows circularly, the allowed regions are located outside the photon sphere, since it is the boundary of circular orbit on which the two conserved quantities $E$ and $L$ diverge. It is predictable that the boundary of the allowed region for circularly moving bulk plasma will approach the photon sphere asymptotically. This feature can also be seen in the right panels of Fig.~\ref{fig:xi-r-p}

For the plunging bulk plasma, within the photon sphere, there are still broad regions allowed for successful energy extraction. The increase in the orientation angle changes the shapes of the allowed regions, enabling them to cover the regions with relatively small black hole spins ($a\lesssim 0.7$), as shown in the left panels of Fig.~\ref{fig:a-r-p}. The event horizon is a natural boundary of allowed regions for energy extraction. Unlike the photon sphere, which serves as the boundary for the allowed region generated by the divergence of circularly flowing bulk plasma, the event horizon acts as a cutoff boundary for the plunging bulk plasma. Hence, it is also a cutoff boundary for the allowed region. This feature can also be observed in the left panels of Fig.~\ref{fig:xi-r-p}

From the right panels of Fig.~\ref{fig:xi-r-p}, it is clear that the allowed regions for energy extraction in the $\xi_{\rm B}–r_0$ planes for circularly flowing bulk plasma are symmetric along the line $\xi_B=0$, as the best orientation angle for circularly flowing bulk plasma is zero (see Eq.\eqref{eq:xi_m} for reference). For the plunging bulk plasma, in the plunging region (the region within the ISCO), the allowed regions for energy extraction in $\xi_{\rm B}–r_0$ planes incline toward the region with positive orientation angles. Additionally, as shown in the left panels of Fig.~\ref{fig:xi-r-p}, this inclination follows the line of the best orientation angle.

\newpage
\bibliographystyle{utphys}
\bibliography{references}

@article{Liu2017,
    author = "Liu, Yi-Hsin and Hesse, M. and Guo, F. and Daughton, W. and Li, H. and Cassak, P. A. and Shay, M. A.",
    title = "{Why does steady-state magnetic reconnection have a maximum local rate of order 0.1?}",
    eprint = "1611.07859",
    archivePrefix = "arXiv",
    primaryClass = "physics.plasm-ph",
    doi = "10.1103/PhysRevLett.118.085101",
    journal = "Phys. Rev. Lett.",
    volume = "118",
    number = "8",
    pages = "085101",
    year = "2017"
}

@article{Lyubarsky2006,
    author = "Lyubarsky, Yury E.",
    title = "{On the relativistic magnetic reconnection}",
    eprint = "astro-ph/0501392",
    archivePrefix = "arXiv",
    doi = "10.1111/j.1365-2966.2005.08767.x",
    journal = "Mon. Not. Roy. Astron. Soc.",
    volume = "358",
    pages = "113--119",
    year = "2005"
}

@article{Comisso:2014nva,
    author = "Comisso, Luca and Asenjo, Felipe A.",
    title = "{Thermal-inertial effects on magnetic reconnection in relativistic pair plasmas}",
    eprint = "1402.1115",
    archivePrefix = "arXiv",
    primaryClass = "physics.plasm-ph",
    doi = "10.1103/PhysRevLett.113.045001",
    journal = "Phys. Rev. Lett.",
    volume = "113",
    pages = "045001",
    year = "2014"
}

@article{KA2008,
    author = "Koide, Shinji and Arai, Kenzo",
    title = "{Energy Extraction from a Rotating Black Hole by Magnetic Reconnection in Ergosphere}",
    eprint = "0805.0044",
    archivePrefix = "arXiv",
    primaryClass = "astro-ph",
    doi = "10.1086/589497",
    journal = "Astrophys. J.",
    volume = "682",
    pages = "1124",
    year = "2008"
}

@article{CA2021,
    author = "Comisso, Luca and Asenjo, Felipe A.",
    title = "{Magnetic Reconnection as a Mechanism for Energy Extraction from Rotating Black Holes}",
    eprint = "2012.00879",
    archivePrefix = "arXiv",
    primaryClass = "astro-ph.HE",
    doi = "10.1103/PhysRevD.103.023014",
    journal = "Phys. Rev. D",
    volume = "103",
    number = "2",
    pages = "023014",
    year = "2021"
}

@article{Yuan2024-1,
    author = "Yang, Hai and Yuan, Feng and Li, Hui and Mizuno, Yosuke and Guo, Fan and Lu, Rusen and Ho, Luis C. and Lin, Xi and Zdziarski, Andrzej A. and Wang, Jieshuang",
    title = "{Modeling the inner part of the jet in M87: Confronting jet morphology with theory}",
    eprint = "2403.15950",
    archivePrefix = "arXiv",
    primaryClass = "astro-ph.HE",
    doi = "10.1126/sciadv.adn3544",
    journal = "Sci. Adv.",
    volume = "10",
    number = "12",
    pages = "adn3544",
    year = "2024"
}

@article{Aimar2023,
    author = "Aimar, N. and Dmytriiev, A. and Vincent, F. H. and Mellah, I. El and Paumard, T. and Perrin, G. and Zech, A.",
    title = "{Magnetic reconnection plasmoid model for Sagittarius A* flares}",
    eprint = "2301.11874",
    archivePrefix = "arXiv",
    primaryClass = "astro-ph.HE",
    doi = "10.1051/0004-6361/202244936",
    journal = "Astron. Astrophys.",
    volume = "672",
    pages = "A62",
    year = "2023"
}

@article{Ripperda2020,
    author = "Ripperda, Bart and Bacchini, Fabio and Philippov, Alexander",
    title = "{Magnetic Reconnection and Hot Spot Formation in Black Hole Accretion Disks}",
    eprint = "2003.04330",
    archivePrefix = "arXiv",
    primaryClass = "astro-ph.HE",
    doi = "10.3847/1538-4357/ababab",
    journal = "Astrophys. J.",
    volume = "900",
    number = "2",
    pages = "100",
    year = "2020"
}

@article{Yuan2009,
    author = "Yuan, Feng and Lin, Jun and Wu, Kinwah and Ho, Luis C.",
    title = "{A Magnetohydrodynamic Model for the Formation of Episodic Jets}",
    eprint = "0811.2893",
    archivePrefix = "arXiv",
    primaryClass = "astro-ph",
    doi = "10.1111/j.1365-2966.2009.14673.x",
    journal = "Mon. Not. Roy. Astron. Soc.",
    volume = "395",
    pages = "2183--2188",
    year = "2009"
}

@article{Yuan2024-2,
    author = "Xi, Lin and Feng, Yuan",
    title = "{Revisiting flares in Sagittarius A* based on general relativistic magnetohydrodynamic numerical simulations of black hole accretion}",
    eprint = "2405.17408",
    archivePrefix = "arXiv",
    primaryClass = "astro-ph.HE",
    doi = "10.1093/mnras/stae1357",
    journal = "Mon. Not. Roy. Astron. Soc.",
    volume = "531",
    number = "3",
    pages = "3136--3150",
    year = "2024"
}

@article{Carleo:2022qlv,
    author = "Carleo, Amodio and Lambiase, Gaetano and Mastrototaro, Leonardo",
    title = "{Energy extraction via magnetic reconnection in Lorentz breaking Kerr\textendash{}Sen and Kiselev black holes}",
    eprint = "2206.12988",
    archivePrefix = "arXiv",
    primaryClass = "gr-qc",
    doi = "10.1140/epjc/s10052-022-10751-w",
    journal = "Eur. Phys. J. C",
    volume = "82",
    number = "9",
    pages = "776",
    year = "2022"
}

@article{Khodadi:2022dff,
    author = "Khodadi, Mohsen",
    title = "{Magnetic reconnection and energy extraction from a spinning black hole with broken Lorentz symmetry}",
    eprint = "2201.02765",
    archivePrefix = "arXiv",
    primaryClass = "gr-qc",
    doi = "10.1103/PhysRevD.105.023025",
    journal = "Phys. Rev. D",
    volume = "105",
    number = "2",
    pages = "023025",
    year = "2022"
}

@article{Wei:2022jbi,
    author = "Wei, Shao-Wen and Wang, Hui-Min and Zhang, Yu-Peng and Liu, Yu-Xiao",
    title = "{Effects of tidal charge on magnetic reconnection and energy extraction from spinning braneworld black hole}",
    eprint = "2201.12729",
    archivePrefix = "arXiv",
    primaryClass = "gr-qc",
    doi = "10.1088/1475-7516/2022/04/050",
    journal = "JCAP",
    volume = "04",
    number = "04",
    pages = "050",
    year = "2022"
}

@article{Liu:2022qnr,
    author = "Liu, Wenshuai",
    title = "{Energy Extraction via Magnetic Reconnection in the Ergosphere of a Rotating Non-Kerr Black Hole}",
    eprint = "2204.07338",
    archivePrefix = "arXiv",
    primaryClass = "astro-ph.HE",
    doi = "10.3847/1538-4357/ac3de3",
    journal = "Astrophys. J.",
    volume = "925",
    number = "2",
    pages = "149",
    year = "2022"
}

@article{Wang:2022qmg,
    author = "Wang, Chao-Hui and Pang, Cheng-Qun and Wei, Shao-Wen",
    title = "{Extracting energy via magnetic reconnection from Kerr\textendash{}de Sitter black holes}",
    eprint = "2209.08837",
    archivePrefix = "arXiv",
    primaryClass = "gr-qc",
    doi = "10.1103/PhysRevD.106.124050",
    journal = "Phys. Rev. D",
    volume = "106",
    number = "12",
    pages = "124050",
    year = "2022"
}

@article{ZhangShaoJun2024,
    author = "Zhang, Shao-Jun",
    title = "{Energy extraction via magnetic reconnection in magnetized black holes}",
    eprint = "2405.16941",
    archivePrefix = "arXiv",
    primaryClass = "gr-qc",
    doi = "10.1088/1475-7516/2024/07/042",
    journal = "JCAP",
    volume = "07",
    pages = "042",
    year = "2024"
}

@article{Shaymatov:2023dtt,
    author = "Shaymatov, Sanjar and Alloqulov, Mirzabek and Ahmedov, Bobomurat and Wang, Anzhong",
    title = "{Kerr-Newman-modified-gravity black hole\textquoteright{}s impact on the magnetic reconnection}",
    eprint = "2307.03012",
    archivePrefix = "arXiv",
    primaryClass = "gr-qc",
    doi = "10.1103/PhysRevD.110.044005",
    journal = "Phys. Rev. D",
    volume = "110",
    number = "4",
    pages = "044005",
    year = "2024"
}

@article{Long:2024tws,
    author = "Long, Fen and Wang, Shangyun and Chen, Songbai and Jing, Jiliang",
    title = "{Magnetic reconnection and energy extraction from a Konoplya{\textendash}Zhidenko rotating non-Kerr black hole}",
    eprint = "2409.11942",
    archivePrefix = "arXiv",
    primaryClass = "gr-qc",
    doi = "10.1140/epjc/s10052-025-13746-5",
    journal = "Eur. Phys. J. C",
    volume = "85",
    number = "1",
    pages = "26",
    year = "2025"
}

@article{Zeng:2025vjt,
    author = "Zeng, Xiao-Xiong and Wang, Ke",
    title = "{Energy Extraction via Magnetic Reconnection in Kerr-Sen-AdS$_{4}$ Black Hole: Circular Plasma and Plunging Plasma}",
    eprint = "2507.10520",
    archivePrefix = "arXiv",
    primaryClass = "gr-qc",
    month = "7",
    year = "2025"
}

@article{Zeng:2025olq,
    author = "Zeng, Xiao-Xiong and Wang, Ke",
    title = "{Energy extraction from the Kerr-Bertotti-Robinson black hole via magnetic reconnection in a circular and a plunging plasma}",
    eprint = "2507.21777",
    archivePrefix = "arXiv",
    primaryClass = "gr-qc",
    doi = "10.1103/vc96-snjm",
    journal = "Phys. Rev. D",
    volume = "112",
    number = "6",
    pages = "064032",
    year = "2025"
}

@article{Work1,
    author = "Shen, Ye and YuChih, Ho-Yun and Chen, Bin",
    title = "{Energy extraction from a rotating black hole via magnetic reconnection: The plunging bulk plasma and orientation angle}",
    eprint = "2409.07345",
    archivePrefix = "arXiv",
    primaryClass = "gr-qc",
    doi = "10.1103/PhysRevD.110.123010",
    journal = "Phys. Rev. D",
    volume = "110",
    number = "12",
    pages = "123010",
    year = "2024"
}

@article{Work0,
    author = "Chen, Bin and Hou, Yehui and Li, Junyi and Shen, Ye",
    title = "{Energy extraction from a Kerr black hole via magnetic reconnection within the plunging region}",
    eprint = "2405.11488",
    archivePrefix = "arXiv",
    primaryClass = "gr-qc",
    doi = "10.1103/PhysRevD.110.063003",
    journal = "Phys. Rev. D",
    volume = "110",
    number = "6",
    pages = "063003",
    year = "2024"
}

@article{Work2,
    title = {Energy extraction from a rotating black hole via magnetic reconnection: Parameters in reconnection models},
    author = {Shen, Ye and YuChih, Ho-Yun},
    eprint = "2412.03010",
    archivePrefix = "arXiv",
    primaryClass = "astro-ph.HE",
    journal = {Phys. Rev. D},
    volume = {111},
    pages = {023003},
    number = "2",
    year = {2025},
    doi = {10.1103/PhysRevD.111.023003},
}

@article{Nathanail:2024efu,
    author = "Nathanail, Antonios and Mizuno, Yosuke and Contopoulos, Ioannis and Fromm, Christian M. and Cruz-Osorio, Alejandro and Moriyama, Kotaro and Rezzolla, Luciano",
    title = "{The impact of resistivity on the variability of black hole accretion flows}",
    eprint = "2411.16684",
    archivePrefix = "arXiv",
    primaryClass = "astro-ph.HE",
    doi = "10.1051/0004-6361/202451836",
    journal = "Astron. Astrophys.",
    volume = "693",
    pages = "A56",
    year = "2025"
}

@article{Penrose,
    author = "Penrose, R.",
    title = "{Gravitational collapse: The role of general relativity}",
    doi = "10.1023/A:1016578408204",
    journal = "Riv. Nuovo Cim.",
    volume = "1",
    pages = "252--276",
    year = "1969"
}

@article{Camilloni:2024tny,
    author = "Camilloni, Filippo and Rezzolla, Luciano",
    title = "{Self-consistent Multidimensional Penrose Process Driven by Magnetic Reconnection}",
    eprint = "2411.04184",
    archivePrefix = "arXiv",
    primaryClass = "gr-qc",
    doi = "10.3847/2041-8213/adbbef",
    journal = "Astrophys. J. Lett.",
    volume = "982",
    number = "1",
    pages = "L31",
    year = "2025"
}

@article{Wald:1974kya,
    author = "Wald, Robert M.",
    title = "{Energy Limits on the Penrose Process}",
    doi = "10.1086/152959",
    journal = "Astrophys. J.",
    volume = "191",
    pages = "231",
    year = "1974"
}

@article{Split-Mono,
    author = {{Michel}, F. Curtis},
    title = "{Rotating Magnetospheres: an Exact 3-D Solution}",
    journal = {Astrophys. J.},
    year = 1973,
    month = mar,
    volume = {180},
    pages = {L133},
    doi = {10.1086/181169},
    adsurl = {https://ui.adsabs.harvard.edu/abs/1973ApJ...180L.133M}
}

@article{Gralla:2014yja,
    author = "Gralla, Samuel E. and Jacobson, Ted",
    title = "{Spacetime approach to force-free magnetospheres}",
    eprint = "1401.6159",
    archivePrefix = "arXiv",
    primaryClass = "astro-ph.HE",
    doi = "10.1093/mnras/stu1690",
    journal = "Mon. Not. Roy. Astron. Soc.",
    volume = "445",
    number = "3",
    pages = "2500--2534",
    year = "2014",
    note = "[Erratum: Mon.Not.Roy.Astron.Soc. 534, 1541 (2024)]"
}

@article{McKinney:2006sc,
    author = "McKinney, Jonathan C.",
    title = "{General relativistic force-free electrodynamics: a new code and applications to black hole magnetospheres}",
    eprint = "astro-ph/0601410",
    archivePrefix = "arXiv",
    doi = "10.1111/j.1365-2966.2006.10087.x",
    journal = "Mon. Not. Roy. Astron. Soc.",
    volume = "367",
    pages = "1797--1807",
    year = "2006"
}

@article{Sasha2011,
    author = {Tchekhovskoy, Alexander and Narayan, Ramesh and McKinney, Jonathan C.},
    title = {Efficient generation of jets from magnetically arrested accretion on a rapidly spinning black hole},
    journal = {Mon. Not. Roy. Astron. Soc. Lett.},
    volume = {418},
    number = {1},
    pages = {L79-L83},
    year = {2011},
    month = {11},
    issn = {1745-3925},
    doi = {10.1111/j.1745-3933.2011.01147.x},
    url = {https://doi.org/10.1111/j.1745-3933.2011.01147.x},
    eprint = {https://academic.oup.com/mnrasl/article-pdf/418/1/L79/56939896/mnrasl\_418\_1\_l79.pdf},
}

@article{Vos:2024loa,
    author = "Vos, Jesse and Cerutti, Benoit and Moscibrodzka, Monika and Parfrey, Kyle",
    title = "{Particle Acceleration in Collisionless Magnetically Arrested Disks}",
    eprint = "2410.19061",
    archivePrefix = "arXiv",
    primaryClass = "astro-ph.HE",
    doi = "10.1103/5wwb-7t2n",
    journal = "Phys. Rev. Lett.",
    volume = "135",
    number = "1",
    pages = "015201",
    year = "2025"
}

@article{Comisso:2016pyg,
    author = "Comisso, L. and Lingam, M. and Huang, Y. -M. and Bhattacharjee, A.",
    title = "{General Theory of the Plasmoid Instability}",
    eprint = "1608.04692",
    archivePrefix = "arXiv",
    primaryClass = "physics.plasm-ph",
    doi = "10.1063/1.4964481",
    journal = "Physics of Plasmas",
    volume = "23",
    pages = "100702",
    year = "2016"
}

@article{Comisso:2017arh,
    author = "Comisso, Luca and Lingam, Manasvi and Huang, Yi-Min and Bhattacharjee, Amitava",
    title = "{Plasmoid Instability in Forming Current Sheets}",
    eprint = "1707.01862",
    archivePrefix = "arXiv",
    primaryClass = "astro-ph.HE",
    doi = "10.3847/1538-4357/aa9789",
    journal = "Astrophys. J.",
    volume = "850",
    number = "2",
    pages = "142",
    year = "2017"
}

@article{MacDonald:1982zz,
    author = "MacDonald, D. and Thorne, K. S.",
    title = "{Black-hole electrodynamics - an absolute-space/universal-time formulation}",
    journal = "Mon. Not. Roy. Astron. Soc.",
    volume = "198",
    pages = "345--383",
    year = "1982"
}

@article{tearing-mode,
    author = {{Furth}, Harold P. and {Killeen}, John and {Rosenbluth}, Marshall N.},
    title = "{Finite-Resistivity Instabilities of a Sheet Pinch}",
    journal = {Physics of Fluids},
    year = 1963,
    month = apr,
    volume = {6},
    number = {4},
    pages = {459-484},
    doi = {10.1063/1.1706761},
    adsurl = {https://ui.adsabs.harvard.edu/abs/1963PhFl....6..459F}
}

@article{Sironi:2025kgn,
    author = "Sironi, Lorenzo and Uzdensky, Dmitri A. and Giannios, Dimitrios",
    title = "{Relativistic Magnetic Reconnection in Astrophysical Plasmas: A Powerful Mechanism of Nonthermal Emission}",
    eprint = "2506.02101",
    archivePrefix = "arXiv",
    primaryClass = "astro-ph.HE",
    doi = "10.1146/annurev-astro-020325-115713",
    month = "6",
    year = "2025"
}

@article{Ebrahimi:2016urr,
    author = "Ebrahimi, F.",
    title = "{Dynamo-driven plasmoid formation from a current-sheet instability}",
    eprint = "1610.09050",
    archivePrefix = "arXiv",
    primaryClass = "physics.plasm-ph",
    doi = "10.1063/1.4972218",
    journal = "Phys. Plasmas",
    volume = "23",
    pages = "120705",
    year = "2016"
}

@article{Tsung:2025eku,
    author = "Tsung, Tsun Hin Navin and Werner, Gregory R. and Uzdensky, Dmitri A. and Begelman, Mitchell C.",
    title = "{Dissipation and particle acceleration in astrophysical jets with velocity and magnetic shear: Interaction of Kelvin-Helmholtz and Drift-Kink Instabilities}",
    eprint = "2501.04090",
    archivePrefix = "arXiv",
    primaryClass = "astro-ph.HE",
    month = "1",
    year = "2025"
}

@article{Sen:1992ua,
    author = "Sen, Ashoke",
    title = "{Rotating charged black hole solution in heterotic string theory}",
    eprint = "hep-th/9204046",
    archivePrefix = "arXiv",
    reportNumber = "TIFR-TH-92-20",
    doi = "10.1103/PhysRevLett.69.1006",
    journal = "Phys. Rev. Lett.",
    volume = "69",
    pages = "1006--1009",
    year = "1992"
}

@article{Casana:2017jkc,
    author = "Casana, R. and Cavalcante, A. and Poulis, F. P. and Santos, E. B.",
    title = "{Exact Schwarzschild-like solution in a bumblebee gravity model}",
    eprint = "1711.02273",
    archivePrefix = "arXiv",
    primaryClass = "gr-qc",
    doi = "10.1103/PhysRevD.97.104001",
    journal = "Phys. Rev. D",
    volume = "97",
    number = "10",
    pages = "104001",
    year = "2018"
}

@article{Jha:2020pvk,
    author = "Jha, Sohan Kumar and Rahaman, Anisur",
    title = "{Bumblebee gravity with a Kerr-Sen-like solution and its Shadow}",
    eprint = "2011.14916",
    archivePrefix = "arXiv",
    primaryClass = "gr-qc",
    doi = "10.1140/epjc/s10052-021-09132-6",
    journal = "Eur. Phys. J. C",
    volume = "81",
    number = "4",
    pages = "345",
    year = "2021"
}

@article{Xu:2022frb,
    author = "Xu, Rui and Liang, Dicong and Shao, Lijing",
    title = "{Static spherical vacuum solutions in the bumblebee gravity model}",
    eprint = "2209.02209",
    archivePrefix = "arXiv",
    primaryClass = "gr-qc",
    doi = "10.1103/PhysRevD.107.024011",
    journal = "Phys. Rev. D",
    volume = "107",
    number = "2",
    pages = "024011",
    year = "2023"
}

@article{Ding:2019mal,
    author = "Ding, Chikun and Liu, Changqing and Casana, R. and Cavalcante, A.",
    title = "{Exact Kerr-like solution and its shadow in a gravity model with spontaneous Lorentz symmetry breaking}",
    eprint = "1910.02674",
    archivePrefix = "arXiv",
    primaryClass = "gr-qc",
    doi = "10.1140/epjc/s10052-020-7743-y",
    journal = "Eur. Phys. J. C",
    volume = "80",
    number = "3",
    pages = "178",
    year = "2020"
}

@article{Panotopoulos:2024jtn,
    author = {Panotopoulos, Grigoris and {\"O}vg{\"u}n, Ali},
    title = "{Strange quark stars and condensate dark stars in Bumblebee gravity}",
    eprint = "2409.05801",
    archivePrefix = "arXiv",
    primaryClass = "gr-qc",
    doi = "10.1016/j.nuclphysb.2025.116956",
    journal = "Nucl. Phys. B",
    volume = "1017",
    pages = "116956",
    year = "2025"
}

@article{Li:2023wlo,
    author = "Li, Chengyi and Ma, Bo-Qiang",
    title = "{Lorentz and CPT breaking in gamma-ray burst neutrinos from string theory}",
    eprint = "2303.04765",
    archivePrefix = "arXiv",
    primaryClass = "hep-ph",
    doi = "10.1007/JHEP03(2023)230",
    journal = "JHEP",
    volume = "03",
    pages = "230",
    year = "2023"
}

@article{Khodadi:2023yiw,
    author = "Khodadi, Mohsen and Lambiase, Gaetano and Mastrototaro, Leonardo",
    title = "{Spontaneous Lorentz symmetry breaking effects on GRBs jets arising from neutrino pair annihilation process near a black hole}",
    eprint = "2302.14200",
    archivePrefix = "arXiv",
    primaryClass = "hep-ph",
    doi = "10.1140/epjc/s10052-023-11369-2",
    journal = "Eur. Phys. J. C",
    volume = "83",
    number = "3",
    pages = "239",
    year = "2023"
}

@article{Ji:2024aeg,
    author = "Ji, Peixiang and Li, Zhuhai and Yang, Lirui and Xu, Rui and Hu, Zexin and Shao, Lijing",
    title = "{Neutron stars in the bumblebee theory of gravity}",
    eprint = "2409.04805",
    archivePrefix = "arXiv",
    primaryClass = "gr-qc",
    doi = "10.1103/PhysRevD.110.104057",
    journal = "Phys. Rev. D",
    volume = "110",
    number = "10",
    pages = "104057",
    year = "2024"
}

@article{Zhu:2024qcm,
    author = "Zhu, Xincheng and Xu, Rui and Xu, Dandan",
    title = "{Bumblebee cosmology: Tests using distance- and time-redshift probes}",
    eprint = "2411.18559",
    archivePrefix = "arXiv",
    primaryClass = "astro-ph.CO",
    month = "11",
    year = "2024"
}

@article{Jha:2021eww,
    author = "Jha, Sohan Kumar and Aziz, Sahazada and Rahaman, Anisur",
    title = "{Study of Einstein-bumblebee gravity with Kerr-Sen-like solution in the presence of a dispersive medium}",
    eprint = "2103.17021",
    archivePrefix = "arXiv",
    primaryClass = "gr-qc",
    doi = "10.1140/epjc/s10052-022-10042-4",
    journal = "Eur. Phys. J. C",
    volume = "82",
    number = "2",
    pages = "106",
    year = "2022"
}

@article{Wang:2021gtd,
    author = "Wang, Zejun and Chen, Songbai and Jing, Jiliang",
    title = "{Constraint on parameters of a rotating black hole in Einstein-bumblebee theory by quasi-periodic oscillations}",
    eprint = "2112.02895",
    archivePrefix = "arXiv",
    primaryClass = "gr-qc",
    doi = "10.1140/epjc/s10052-022-10475-x",
    journal = "Eur. Phys. J. C",
    volume = "82",
    number = "6",
    pages = "528",
    year = "2022"
}

@article{Islam:2024sph,
    author = "Islam, Shafqat Ul and Ghosh, Sushant G. and Maharaj, Sunil D.",
    title = "{Investigating rotating black holes in bumblebee gravity: insights from EHT observations}",
    eprint = "2410.05395",
    archivePrefix = "arXiv",
    primaryClass = "gr-qc",
    doi = "10.1088/1475-7516/2024/12/047",
    journal = "JCAP",
    volume = "12",
    pages = "047",
    year = "2024"
}

@article{Gu:2022grg,
    author = "Gu, Jiale and Riaz, Shafqat and Abdikamalov, Askar B. and Ayzenberg, Dimitry and Bambi, Cosimo",
    title = "{Probing bumblebee gravity with black hole X-ray data}",
    eprint = "2206.14733",
    archivePrefix = "arXiv",
    primaryClass = "gr-qc",
    doi = "10.1140/epjc/s10052-022-10686-2",
    journal = "Eur. Phys. J. C",
    volume = "82",
    number = "8",
    pages = "708",
    year = "2022"
}

@article{Mai:2024lgk,
    author = "Mai, Zhan-Feng and Xu, Rui and Liang, Dicong and Shao, Lijing",
    title = "{Dynamic instability analysis for bumblebee black holes: The odd parity}",
    eprint = "2401.07757",
    archivePrefix = "arXiv",
    primaryClass = "gr-qc",
    doi = "10.1103/PhysRevD.109.084076",
    journal = "Phys. Rev. D",
    volume = "109",
    number = "8",
    pages = "084076",
    year = "2024"
}

@article{Kostelecky:2003fs,
    author = "Kostelecky, V. Alan",
    title = "{Gravity, Lorentz violation, and the standard model}",
    eprint = "hep-th/0312310",
    archivePrefix = "arXiv",
    reportNumber = "IUHET-461",
    doi = "10.1103/PhysRevD.69.105009",
    journal = "Phys. Rev. D",
    volume = "69",
    pages = "105009",
    year = "2004"
}

@article{Bailey:2006fd,
    author = "Bailey, Quentin G. and Kostelecky, V. Alan",
    title = "{Signals for Lorentz violation in post-Newtonian gravity}",
    eprint = "gr-qc/0603030",
    archivePrefix = "arXiv",
    reportNumber = "IUHET-489",
    doi = "10.1103/PhysRevD.74.045001",
    journal = "Phys. Rev. D",
    volume = "74",
    pages = "045001",
    year = "2006"
}

@article{Sen:2025ytv,
    author = "Sen, Gargi and Maity, Debaprasad and Das, Santabrata",
    title = "{GRMHD modelling of accretion flow around Sagittarius A* constrained by EHT measurements}",
    eprint = "2510.03602",
    archivePrefix = "arXiv",
    primaryClass = "astro-ph.HE",
    month = "10",
    year = "2025"
}

\end{document}